\documentclass[journal,twoside,web]{ieeecolor}
\usepackage{comment}
\usepackage{generic}
\usepackage{cite}
\usepackage{amsmath,amssymb,amsfonts}
\usepackage[dvipsnames]{xcolor}
\usepackage{hhline,colortbl,arydshln}
\usepackage{algorithm,algorithmic}
\usepackage{graphicx}
\usepackage{hyperref}
\hypersetup{hidelinks=true}
\usepackage{textcomp}
\usepackage{tabularx,multirow}
\usepackage{diagbox}
\usepackage{soul}
\usepackage{booktabs}
\usepackage{scalerel}

\def\BibTeX{{\rm B\kern-.05em{\sc i\kern-.025em b}\kern-.08em
    T\kern-.1667em\lower.7ex\hbox{E}\kern-.125emX}}
    
\newcolumntype{Y}{>{\centering\arraybackslash}X}
\newlength{\charlength}
\begin{document}
\title{TrustEMG-Net: Using Representation-Masking Transformer with U-Net for Surface Electromyography Enhancement}
\author{Kuan-Chen Wang, Kai-Chun Liu, Ping-Cheng Yeh, Sheng-Yu Peng, and Yu Tsao, \IEEEmembership{Senior Member, IEEE}
\thanks{This work was supported in part by National Science and Technology Council of Taiwan and Academia Sinica. (Corresponding Authors: Kai-Chun Liu; Yu Tsao)}
\thanks{Kuan-Chen Wang is with the Graduate Institute of Communication Engineering, National Taiwan University, Taiwan (email: d12942016@ntu.edu.tw). }
\thanks{Kai-Chun Liu is with the College of Information and
Computer Sciences, University of Massachusetts Amherst, MA, 01003, USA (email: kaichunliu@umass.edu).}
\thanks{Ping-Cheng Yeh is with the Graduate Institute of Communication Engineering, National Taiwan University, Taiwan (email: pcyeh@ntu.edu.tw).}
\thanks{Sheng-Yu Peng is with the Department of Electrical Engineering, National Taiwan University of Science and Technology, Taiwan (email: sypeng@mail.ntust.edu.tw)}
\thanks{Yu Tsao is with the Research Center for Information Technology Innovation,
Academia Sinica, Taiwan, and also with the Department of Electrical
Engineering, Chung Yuan Christian University, Taiwan (email: yu.tsao@citi.sinica.edu.tw).}
}

\maketitle

\begin{abstract}
Surface electromyography (sEMG) is a widely employed bio-signal that captures human muscle activity via electrodes placed on the skin. Several studies have proposed methods to remove sEMG contaminants, as non-invasive measurements render sEMG susceptible to various contaminants. However, these approaches often rely on heuristic-based optimization and are sensitive to the contaminant type. A more potent, robust, and generalized sEMG denoising approach should be developed for various healthcare and human\textendash computer interaction applications. This paper proposes a novel neural network (NN)-based sEMG denoising method called TrustEMG-Net. It leverages the potent nonlinear mapping capability and data-driven nature of NNs. TrustEMG-Net adopts a denoising autoencoder structure by combining U-Net with a Transformer encoder using a representation-masking approach. The proposed approach is evaluated using the Ninapro sEMG database with five common contamination types and signal-to-noise ratio (SNR) conditions. Compared with existing sEMG denoising methods, TrustEMG-Net achieves exceptional performance across the five evaluation metrics, exhibiting a minimum improvement of 20\%. Its superiority is consistent under various conditions, including SNRs ranging from -14 to 2 dB and five contaminant types. An ablation study further proves that the design of TrustEMG-Net contributes to its optimality, providing high-quality sEMG and serving as an effective, robust, and generalized denoising solution for sEMG applications.
\end{abstract}

\begin{IEEEkeywords}
Surface electromyography, sEMG contaminant removal, denoising autoencoder, U-Net, Transformer.
\end{IEEEkeywords}

\section{Introduction}
\label{sec:introduction}
Surface electromyography (sEMG) is a bio-signal that monitors human muscle activity. During muscle contraction, the electric potential changes generated by motor units are captured by electrodes placed on the skin. The non-invasive nature makes sEMG widely adopted in various applications, such as neuromuscular system investigation~\cite{tang2018novel}, respiration monitoring~\cite{silva2023surface}, rehabilitation~\cite{engdahl2015surveying}, human\textendash computer interfaces~\cite{ma2014hand}, and speech aids~\cite{scheck2023multi}. However, its non-invasive measurement also renders sEMG susceptible to various contaminants, such as baseline wander (BW), powerline interference (PLI), electrocardiogram (ECG), motion artifacts (MOA), and white Gaussian noise (WGN)~\cite{boyer2023reducing,farago2022review}. These contaminants distort the amplitude and frequency spectrum of the sEMG, hindering information extraction and interpretation~\cite{farago2022review}. Hence, developing contaminant removal methods for sEMG is crucial to maintaining the efficacy and robustness of sEMG applications.

Several studies have proposed sEMG contaminant removal methods, which are categorized as single-channel or multi-channel approaches. Single-channel sEMG denoising techniques, including infinite impulse response (IIR) filters, template subtraction (TS)~\cite{xu2020comparative,drake2006elimination}, and decomposition-based methods~\cite{ashraf2021threshold,sun2020surface,torres2011complete,ma2020emg}, rely solely on noisy sEMG signals, whereas multi-channel techniques, including adaptive filtering~\cite{marque2005adaptive} and independent component analysis~\cite{willigenburg2012removing}, leverage auxiliary data, such as other sEMG channels or reference ECG or PLI signals. This study focused on developing single-channel sEMG denoising methods that offer flexibility across electrode types (e.g., single pairs or array-type electrodes) and avoid the need to collect and process the auxiliary data.

Existing single-channel sEMG denoising methods have certain limitations, including lack of effectiveness, robustness, and generalizability for a more generalized denoising scenario. For example, IIR filters inevitably discard the frequency components of sEMG; thus, they are unsuitable for removing contaminants with a broad frequency range~\cite{boyer2023reducing}. TS depends on certain assumptions to achieve decent performance in removing quasi-periodic ECG artifacts, such as sEMG following a zero-mean Gaussian distribution or precise ECG segment detection, which may not hold in practical scenarios~\cite{peri2021singular, xu2020comparative}. Methods based on signal decomposition, such as empirical mode decomposition (EMD), ensemble EMD (EEMD), complete EEMD with adaptive noise (CEEMDAN), and variational mode decomposition (VMD), leverage data-driven characteristics during the signal decomposition stage~\cite{ashraf2021threshold,sun2020surface,ma2020emg}. However, the denoising process relies heavily on human expertise and trial and error, resulting in a complex optimization process that is susceptible to suboptimal denoising performance.

Recently, neural networks (NNs) have gained prominence for signal enhancement, such as in acoustics~\cite{lu2013speech, FCN_fu2017raw}, ECG~\cite{chiang2019noise,hu2024lightweight}, and EEG~\cite{zhang2021eegdenoisenet}. In these studies, NN-based noise-removal techniques outperform traditional methods, which is primarily attributed to the effective nonlinear mapping capability and data-driven nature of NNs. Various NN models have been incorporated in these studies, including multilayer perceptrons (MLP)~\cite{lu2013speech,zhang2021eegdenoisenet}, convolutional neural networks (CNNs)~\cite{zhang2021eegdenoisenet,chiang2019noise}, long short-term memory models (LSTM)~\cite{zhang2021eegdenoisenet,LSTM_weninger2015speech}, fully convolutional neural networks (FCNs)~\cite{FCN_fu2017raw, chiang2019noise}, U-Net~\cite{hu2024lightweight,defossez2020real}, and Transformer~\cite{wang2021tstnn}. Several studies have explored NN-based approaches for sEMG noise removal~\cite{kale2009intelligent, wang2023ecg}. However, these studies primarily developed techniques using typical MLP-based models~\cite{kale2009intelligent,mankar2011emg}, or addressed specific noise types (e.g., PLI~\cite{kale2009intelligent} or ECG~\cite{wang2023ecg}). These limitations may constrain the effectiveness and usability of NN-based sEMG denoising. Hence, a more effective and suitable NN model structure for sEMG denoising must be developed, and its robustness and generalizability should be validated for various contaminant types.

To address the limitations of the existing sEMG denoising methods, we propose an NN-based approach called TrustEMG-Net. TrustEMG-Net integrates U-Net with a Transformer encoder layer as a denoising autoencoder (DAE). Previous studies~\cite{oktay2018attention, azad2022contextual} demonstrated that U-Net captures local information excellently through its convolutional layers and residual architecture. Other studies~\cite{azad2022contextual, vaswani2017attention} have shown that Transformers are highly effective in modeling global (long-term) information in sequential signals using a self-attention mechanism. Integration automatically utilizes local and global information within the data to remove contaminants. Furthermore, previous studies~\cite{tsai2020blind, hsu2017unsupervised, hsu2016voice} have confirmed that signals that may combine multiple ingredients can be more easily disentangled and analyzed separately in a high-dimensional representation space compared with the original physical domain. Based on this concept, in this study, we employed the representation-masking (RM) approach for the Transformer encoder to explicitly conduct denoising in a high-level representation space. The experimental results demonstrated that TrustEMG-Net outperforms existing sEMG denoising methods by achieving superior signal quality and low feature extraction errors across various signal-to-noise ratios (SNRs) and contaminant types. The RM approach can also enhance the denoising performance, particularly for narrow-band contaminants. These findings suggest that TrustEMG-Net offers an effective and robust model architecture suitable for sEMG waveform denoising, demonstrating its potential for providing high-quality signals for sEMG applications.

The main contributions of this study are as follows:
\begin{itemize}
  \item We developed an NN-based sEMG denoising method called TrustEMG-Net. TrustEMG-Net leverages the strengths of U-Net and Transformer encoders to capture both local and global information and effectively handle various contaminants. The RM approach utilizes a Transformer encoder to preserve important features for the U-Net's decoder to reconstruct the sEMG signals. An ablation study proved that integrating the U-Net and Transformer encoder is effective for sEMG denoising, and the RM approach further boosts the performance, particularly with narrow-band contaminants.
  \item We conducted a comprehensive experiment to analyze the denoising capabilities of the existing and proposed approaches. We utilized five common contamination types and SNR conditions, as well as five evaluation metrics focusing on signal quality and clinical indices to validate the generalization, effectiveness, and robustness of the denoising techniques. 
  \item The proposed method was rigorously compared with existing sEMG contaminant removal techniques via cross-validation. TrustEMG-Net consistently outperformed existing methods, achieving significantly improved signal quality reconstruction and reduced feature extraction errors. This superiority was observed across a broad range of SNRs and contaminant types.
  \item This study contributes to the development of a more effective, robust, and generalized sEMG contaminant removal method. TrustEMG-Net has the potential to provide high-quality sEMG signals for clinical and human\textendash computer interaction (HCI) applications, where precise interpretation and information extraction are critical in practical practice.\footnote{Once the paper is accepted, the codes would be available at \url{https://github.com/eric-wang135/TrustEMG}.}
\end{itemize}

The remainder of this paper is organized as follows: Section II reviews existing sEMG denoising methods and NN models used in the proposed method. Section III introduces the database and proposed sEMG contaminant removal method. Section IV presents the experimental results. Section V analyzes the results, discusses the limitations, and outlines future work. Finally, Section VI concludes the paper.

\section{Related works}
\subsection{Single-channel sEMG contaminant removal methods}
Several methods have been developed for eliminating contaminants from single-channel sEMG. A classic approach involves using digital filters, particularly IIR filters~\cite{boyer2023reducing,ma2020emg}. These filters can suppress noise in specific frequency bands by selecting appropriate filter types and corresponding parameters, including the cutoff frequencies and quality factors. For instance, high-pass filters with 10\textendash 40 Hz cutoff frequencies are often employed to remove the BW, ECG, and MOA~\cite{boyer2023reducing,ma2020emg,xu2020comparative}. Bandstop filters with central frequencies of 50 (or 60) Hz and their harmonics are suitable for managing PLI~\cite{boyer2023reducing,ma2020emg}, and bandpass filters with cutoff frequencies 10\textendash 500 Hz are commonly used for WGN~\cite{boyer2023reducing,ma2020emg}. IIR filters have certain advantages, such as ease of implementation and computational efficiency. However, they also have limitations such as the removal of specific frequency components of the sEMG, which reduces their effectiveness against noise with broad frequency distributions.

TS is a denoising technique suitable for removing quasi-periodic artifacts such as ECG~\cite{boyer2023reducing,xu2020comparative}. TS involves a three-step process: detection of ECG signal segments using specialized algorithms, creation of templates by averaging or filtering noisy sEMG segments, and the subtraction of these templates from the corresponding sEMG segments~\cite{xu2020comparative,junior2019template}. Unlike IIR filters, TS avoids the removal of specific frequency bands, thus preserving the integrity of the sEMG signal. However, the success of TS relies on certain assumptions, including the sEMG following a zero-mean Gaussian distribution and accurate ECG segment detection, which may not hold in practical scenarios.

Signal decomposition algorithms, including EMD~\cite{huang1998empirical}, EEMD~\cite{wu2009ensemble}, CEEMDAN~\cite{torres2011complete}, and VMD~\cite{dragomiretskiy2013vmd}, have been used in single-channel sEMG denoising. These methods decompose noisy sEMG signals into intrinsic mode functions (IMFs). Subsequently, modes with contaminants are further filtered, thresholded, or discarded based on the contaminant type and denoising techniques~\cite{ashraf2021threshold,sun2020surface,ma2020emg,sun2023emg,ashraf2023variational}. Subsequently, the enhanced sEMG waveform is reconstructed by summing the remaining modes. The primary advantage of decomposition-based methods lies in their data-driven characteristics and the ability to manage broadband and stochastic noise, namely, WGN~\cite{ashraf2021threshold,ashraf2023variational}. However, these methods may struggle to optimize denoising performance. Only the decomposition process benefits from its data-driven nature. In contrast, the denoising process relies heavily on human expertise and trial and error, which can result in suboptimal denoising outcomes.

\subsection{NN-based denoising autoencoder}
A DAE is an autoencoder designed to reconstruct clean signals from noisy ones~\cite{vincent2008extracting}. DAEs consist of encoder and decoder components. The encoder transforms the noisy input signal into a hidden representation, and the decoder maps the hidden representation to the reconstructed signal to closely approximate the ground truth using criteria such as L1 or L2 loss. Several studies have demonstrated the effectiveness of DAEs in reducing noise in various domains, including medical images~\cite{gondara2016medical}, speech~\cite{lu2013speech,defossez2020real}, and bio-signals~\cite{chiang2019noise,li2015feature}.

U-Net is an NN model derived from the FCN architecture that was initially developed for semantic segmentation in medical imaging~\cite{ronneberger2015u}. The vanilla U-Net architecture comprises an encoder and a decoder constructed using convolutional layers. The encoder extracts feature maps from the input data, whereas the decoder uses these feature maps to generate an output that matches the dimension of the original input. Skip connections distinguish U-Net from typical FCNs, where feature maps from the encoder layers are directly passed to the corresponding layers in the decoder. This design aids in producing high-quality output data because the decoder can leverage multi-scale information instead of relying purely on the high-level feature maps of the encoder output, which tend to discard low-level information. Thus, U-Net has also been adapted to implement DAEs in various data types such as speech and biomedical signals~\cite{hu2024lightweight,defossez2020real}.

\subsection{Transformer}
Transformer is effective because it can process sequences in parallel using a self-attention mechanism~\cite{vaswani2017attention}. This characteristic makes Transformer more effective in capturing long-term dependencies and more efficient in the training and inference stages. These advantages make Transformer widely adopted in the signal denoising field~\cite{wang2021tstnn,wang2021caunet}.

Transformer consists of encoder and decoder components, with the encoder being the focus here. The Transformer encoder includes a positional encoding module and multiple encoder layers, comprising multi-head attention, layer normalization, and a feedforward neural network. Residual connections are applied to address the vanishing gradient.

The positional encoding module embeds the sequence order information into the data, enabling the model to consider the data content and position. This can be achieved through parameterized embedding layers or fixed functions such as sinusoidal functions. Subsequently, the multi-head self-attention module captures the relationships among the data sequences. The self-attention mechanism involves query, key, and value vectors, and the attention weights are computed using the scaled dot-product and softmax function, as shown in Eq.~\ref{eq:Attention}.
\begin{equation}
\label{eq:Attention}
    \text{Attention}\{Q,K,V\} = \text{softmax}(\frac{QK^T}{\sqrt{d_h}})V,
\end{equation}
where $Q$, $K$, and $V$ denote query, key, and value in matrix form, respectively, and $d_h$ represents the embedding dimension designed to prevent numerical instability resulting from large vector dimensions.

The outputs from the multiple self-attention heads are concatenated and transformed to match the dimensions of the input, as shown in Eqs.~\ref{eq:Multihead-Attention} and~\ref{eq:Multihead-Attention-head}. 
\begin{equation}
\label{eq:Multihead-Attention}
    \text{Multihead}(Q,K,V) = \text{concat}(head_1,...,head_h)W^O,    
\end{equation}
\begin{equation}
\label{eq:Multihead-Attention-head}
    head_i = \text{Attention}(QW_i^Q,KW_i^K,VW_i^V),
\end{equation}
where $h$ denotes the number of attention heads, and $W_i^Q$, $W_i^K$, $W_i^V$, and $W^O$ represent the parameters of the projection matrices. The output of the multi-head attention module is added to the input through residual connections and normalized using layer normalization, making it suitable for sequences of varying lengths.

\section{Methodology}
\subsection{Dataset and preprocessing}
\subsubsection{Clean sEMG database}

This study utilized the Non-Invasive Adaptive Prosthetics (Ninapro) database as a clean sEMG data source~\cite{atzori2014electromyography}. Twelve sEMG channels were recorded using active wireless electrodes placed on the upper arm, with a sampling rate of 2 kHz. The Ninapro database contains three subsets of data: DB1, DB2, and DB3. In this study, we adopted DB2, which comprises sEMG data collected from 40 intact subjects and thus has the most extensive subject pool. DB2 has three sessions: Exercises 1, 2, and 3. This study used the first two sessions, in which the subjects performed 17 and 22 types of movements, respectively. Each movement type was executed six consecutive times, with a duration of five seconds, followed by a three-second rest period. Notably, previous studies have viewed this database as a reliable source of clean sEMG data after applying appropriate filtering techniques~\cite{machado2021deep,wang2023ecg,zhang2023semg}. We omitted Ninapro DB1 and DB3 as they are unsuitable for clean sEMG datasets. DB1 only provides the rectified version of the sEMG signals and not the raw sEMG signals~\cite{atzori2014electromyography}. As our study focused on denoising raw sEMG signals, DB1 was inappropriate for our purposes. DB3 includes data from 11 transradial amputees, which is a relatively small and unique subject pool. Moreover, DB3 potentially contains more contamination and may not be a reliable source of clean sEMG data~\cite{machado2021deep}.

We initiated a series of preprocessing steps to prepare the clean sEMG data. The sEMG data were passed through a fourth-order Butterworth bandpass filter with cutoff frequencies of 20 Hz and 500 Hz. Subsequently, the sEMG signals were downsampled to 1 kHz and normalized by dividing them by the maximum absolute value. Finally, we segmented the sEMG signals into 2-s segments and removed the silent ones. Note that although the preprocessing discards sEMG components below 20 Hz and above 500 Hz, this study aimed to restore the main frequency components of sEMG within the 20\textendash 500 Hz frequency band. This range contains the most useful information and is commonly employed in various sEMG applications~\cite{xiaojing2011feature, thongpanja2013mean, cao2022control}.

\subsubsection{sEMG contaminants dataset}
Five types of sEMG contaminants were used. The sampling rates of these contaminants were unified at 1 kHz to match with the sEMG data. In the following sections, we describe the characteristics of these five contaminants.

\paragraph{BW} 
We sourced data from the MIT-BIH Noise Stress Test Database (NSTDB)~\cite{moody1984noise} to simulate BW. This database includes three types of noise that are commonly encountered in ECG signals: BW, muscle artifacts, and electrode motion artifacts. The BW data in the MIT-BIH NSTDB were adopted as the sEMG contaminant, which is reasonable because both sEMG and ECG signals use electrode patches for measurements~\cite{yadav2023noise}. 

\paragraph{PLI} 
PLI is often simulated using a sinusoidal wave with a frequency of 50 or 60 Hz~\cite{machado2021deep,boyer2023reducing}. To increase the realism of the noise and elevate the denoising challenge, this study incorporated a frequency drift up to 1.5 Hz~\cite{mateo2008neural}. Consequently, the sine wave frequency varied between 58.5 and 61.5 Hz.

\paragraph{ECG} 
We obtained ECG data from the MIT-BIH Normal Sinus Rhythm Database (NSRD), containing 24 h of ECG signals from 14 subjects~\cite{goldberger2000physiobank}. In related research on sEMG denoising and contaminant-type identification, data from the MIT-BIH NSRD are often used as a source of ECG interference in sEMG~\cite{machado2021deep,wang2023ecg}. We applied a sequential preprocessing approach to mitigate the potential influence of noise components in the ECG data, such as BW, PLI, and high-frequency environmental noise. This included a high-pass filter with a cutoff frequency of 1 Hz, a notch filter centered at 60 Hz, and a low-pass filter with a cutoff frequency of 200 Hz~\cite{mccool2014identification,abdelazez2018detection}. 

\paragraph{MOA} 
Two MOA data sources were incorporated, considering their diversity. Reference ~\cite{machado2021deep} describes the first type of MOA. This dataset followed the sEMG collection method of the Ninapro database, which involved the placement of 12 electrodes at the same positions. During a 10-s recording period, the electrodes were lightly tapped every second to generate noise. We applied a 51-point moving average filter to preprocess the dataset, following the previous research~\cite{machado2021deep}. The second type of MOA is the electrode motion artifacts from the MIT-BIH NSTDB~\cite{moody1984noise}, which has been employed in previous studies on the removal of sEMG contaminants~\cite{mccool2014identification}. 


\paragraph{WGN} 
WGN was mathematically simulated using a random function from the NumPy library~\cite{harris2020array}. 

\paragraph{Combined contaminant}
To introduce more complex denoising scenarios, our experiment incorporated compound contaminants generated by combining multiple contaminant types~\cite{ma2020emg, sauer2024signal}, including mixtures of three and five contaminant types. Five-type compound contaminants represent highly complicated denoising scenarios, whereas three-type compound contaminants can emulate more scenarios encountered in sEMG applications, considering that not every contaminant type may be present in every use case.

The three-type compound contaminant has 10 combinations for mixing by selecting three out of five distinct contaminant types. Each contaminant type contributes an equal signal energy to the compound noise. The same principle applies to the mixing of the five types of contaminants.
\subsubsection{Noisy sEMG dataset}
The experimental noisy sEMG data were synthesized by superimposing contaminants onto clean sEMG at different SNRs, which were calculated as follows:

\begin{equation}
\text{SNR} = 10\log_{10}\left(\frac{P_{\text{signal}}}{P_{\text{noise}}}\right) = 10\log_{10}\frac{\sum_{i=1}^{k}x_{s_i}^2}{\sum_{i=1}^{k}x_{n_i}^2},
\end{equation}
where \(P_{\text{signal}}\) and \(P_{\text{noise}}\) represent the power of the clean sEMG and contaminant, respectively; \(x_s\) and \(x_n\) denote the waveforms of the clean sEMG and contaminant, respectively; and \(k\) is the length of the sEMG signal segment. 


\subsection{Data preparation}
Mismatch conditions were introduced between the training and test sets to fairly evaluate the denoising performance of the deep learning models, including the sources of sEMG and contaminants, characteristics of contaminants, and SNRs of noisy signals. Table~\ref{tab:mismatch} summarizes the conditional differences between the training and test sets. Regarding the sEMG data, the training and test sets utilized data from different channels (electrode locations can affect the sEMG signal properties~\cite{reaz2006techniques}), exercise sessions, and subjects within Ninapro database DB2. The training and validation sets consisted of sEMG recordings from Channel 2 during Exercise 1 by 25 and 5 subjects, respectively. The test set included sEMG recordings from Channel 11 during Exercise 2 by 10 subjects. The numbers of clean sEMG segments in the training, validation, and test sets were 10804, 2664, and 3180, respectively.

The sources and characteristics of the sEMG contaminants also differed between the training and test sets. BW and MOA were sourced from different channels, PLI was generated using sine waves with varying frequencies and phases, and ECG artifacts were obtained from different subjects. Additionally, the SNRs of the noisy signals also exhibited differences. The training set featured SNR values of 1, -3, -7, -11, and -15 dB, whereas the test set encompassed SNR values of 2, -2, -6, -10, and -14 dB.

Based on the synthesis conditions outlined earlier, the training, validation, and test sets comprised 324120 (10804$\times$ 5 SNRs$\times$ 6 contamination types), 53280 (2664$\times$ 5 SNRs$\times$ 4 contamination types), and 95400 (3180$\times$ 5 SNRs$\times$ 6 contamination types) noisy sEMG segments. In each dataset, segments featuring one, three, or five types of noise accounted for approximately one-third each, ensuring a balanced representation of each noise type.
\begin{table*}[t]
    \centering
    \caption{Data sources and mismatch conditions between the training and test sets.}
    \label{tab:mismatch}
    \begin{tabular}{ccccc}
     \toprule
     Data type & Data source & {Mismatch Condition} & Training and validation set & Test set\\
     \midrule
    & & Subject & 30 subjects (25 for training, 5 for validation)&  10 subjects \\ 
    sEMG & Ninapro database~\cite{atzori2014electromyography} & Session & Exercise 1	& Exercise 2 \\
    & & Channel &  2 (elbow)	& 11 (biceps brachii)\\
    \midrule
    BW & MIT-BIH NSTDB~\cite{moody1984noise} & Channel &	1	&  2\\
    
    PLI & Simulation & Frequency / Phase & 58.4 to 61.4 Hz (step of 0.2 Hz)& 58.8 to 61.5 Hz (step of 0.375 Hz)\\
    
    ECG & MIT-BIH NSRD~\cite{goldberger2000physiobank} & Subject &	14 subjects & 4 subjects(19090, 19093, 19140, 19830) \\
    \multirow{2}{*}{MOA} & Machado et. al\cite{machado2021deep} & Channel & 1-8 & 9-12\\
    & MIT-BIH NSTDB ~\cite{moody1984noise} & Channel & 1 & 2\\
    WGN & Simulation & - & - & -\\
    \midrule
    - & - & SNR &  1, -3, -7, -11, and -15 dB & 2, -2, -6, -10 and -14 dB \\
    \bottomrule
    \end{tabular}
\end{table*}

\begin{figure}[t!]
    \centering
    \includegraphics[width=\columnwidth]{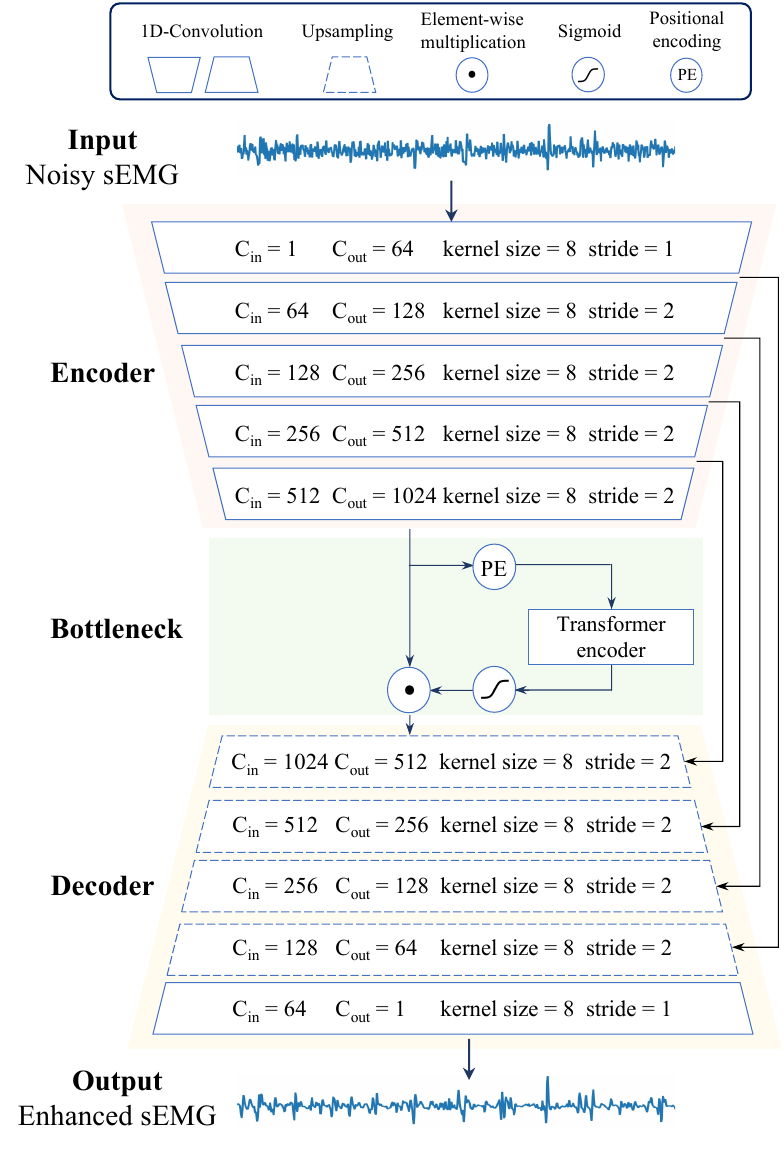}
    \caption{The architecture of the proposed TrustEMG-Net.}
    \label{fig: Model structure}
\end{figure}

\subsection{Proposed method}
The proposed sEMG denoising method, TrustEMG-Net, is shown in Fig.~\ref{fig: Model structure}. TrustEMG-Net is designed as an end-to-end system that uses a noisy sEMG waveform as the input and generates a clean sEMG waveform as the output. Integrating U-Net with a Transformer encoder layer as a DAE enables the model to extract both local and global information for contaminant removal, thereby effectively eliminating various sEMG contaminants. Specifically, the convolution layers of the U-Net can capture local information within a noisy sEMG input~\cite{azad2022contextual}. Local information is associated with the patterns or statistical properties of waveform segments, representing the local characteristics of a signal, such as peaks, crossing points, power, envelopes, and frequency. These patterns or statistical properties provide useful clues for denoising. Thus, several typical denoising methods, such as median filtering~\cite{bacspinar2015comparative} and interval thresholding~\cite{ashraf2023variational}, rely on such local information to remove noise by segment. In contrast, the Transformer encoder excels in capturing global information through its attention mechanism, compensating for the limited receptive field of convolution operations~\cite{vaswani2017attention,azad2022contextual}. Global information is obtained by simultaneously processing the entire input sequence, which may be related to the trend, periodicity, or overall power of the signal or noise. Denoising techniques that use global information, such as wavelet denoising~\cite{sobahi2011denoising} and non-local means~\cite{kumar2018denoising}, can better identify and mitigate contaminants that have a broader or more systemic impact on a signal.

Furthermore, the Transformer encoder employs an RM approach that helps the model learn to denoise at a high-level representation. Masking is a denoising technique often used with generic transformations, such as the short-time Fourier transform~\cite{jiang2011performance,fu2021metricgan+}. A mask can suppress unwanted signals and preserve the target signals by assigning weights close to zero or one to different parts of a representation. Moreover, many studies have shown that the use of learnable transformations (i.e., NN-based encoders and decoders) with masking techniques can further enhance performance in denoising and signal separation~\cite{luo2019conv,kadiouglu2020empirical,li2023u}. Generic transformations may not be optimal for various denoising tasks. In TrustEMG-Net, the U-Net’s encoder and decoder serve as learnable transformations; thus, the Transformer encoder can function as a mask predictor to implement the RM technique.

The architecture of TrustEMG-Net is explained as follows:

\subsubsection{Encoder and decoder}
The U-Net consists of an encoder and a decoder. The encoder comprises five convolutional layers, and the decoder has four upsampling modules and a transposed convolutional layer. The kernels of the convolutional and transposed convolutional layers in the U-net are one-dimensional with a size of 8. Each output from the convolutional layers of the encoder is connected to the corresponding upsampling module in the decoder through skip connections. The upsampling module consists of a transposed convolutional layer (stride of two) and a convolutional layer (stride of one). Except for the final transposed convolutional layer, both the convolutional and transposed convolutional layers in U-Net employ batch normalization and ReLU activation functions. 

Assuming a noisy input signal of dimension $d \times 1$, the encoder transforms the input vector into a latent representation with $\frac{d}{16} \times 1024$ dimensions, where $d$ denotes the number of samples in the input signal. The first convolutional layer with a stride of 1 transforms the input vector into representations with dimensions $d \times 64$. The second to fifth convolutional layers, each with a stride of 2, compress the time dimension of the input by half while expanding the feature dimension by 2. Consequently, the representations are transformed into the dimensions of $\frac{d}{2} \times 128$, $\frac{d}{4} \times 256$, $\frac{d}{8} \times 512$, and $\frac{d}{16} \times 1024$ layer by layer. The decoder then reconstructs the clean sEMG signal from the masked feature maps. Using a symmetric design, the decoder employs four upsampling modules to transform the $\frac{d}{16} \times 1024$ latent representations into $\frac{d}{8} \times 512$, $\frac{d}{4} \times 256$, $\frac{d}{2} \times 128$, and $d \times 64$ dimensions. Finally, the last transposed convolutional layer with a stride of 1 converts the $d \times 64$ representation back to a $d \times 1$-dimensional vector.

\subsubsection{Bottleneck}
In the bottleneck of U-Net, the encoder's output undergoes positional encoding using a sinusoidal function. It is subsequently fed into a single Transformer encoder layer, as shown in Fig.~\ref{fig: Model structure} (b). The Transformer encoder layer has an embedding dimension of 1024, 8 attention heads, a feedforward network dimension of 2048, and a dropout rate of 0.1. The Transformer encoder is applied using the RM approach, which can be expressed as follows:
\begin{equation}
\label{eq:RM}
    \hat{x} = \text{sigmoid}(f(x))\odot x,
\end{equation}
where $x$ denotes the representation of the encoder output, $\hat{x}$ is the representation for the decoder input, $f$ represents the transformation of the Transformer encoder, and $\odot$ denotes the Hadamard product. The RM approach uses the Transformer encoder output as a mask for the latent representation. The output of the Transformer encoder layer passes through a sigmoid activation function to generate a mask with values between 0 and 1 and dimensions identical to the input representations. This mask is then multiplied element-wise by the latent representations from the U-Net encoder, performing a masking operation. Finally, the masked representations are passed to the decoder.

\subsection{Implementation details}
In this study, four-fold subject-wise cross-validation was conducted four times, with each fold selecting 10 out of 40 subjects for testing. We trained the models using the Adam optimizer~\cite{kinga2015method} and L1 loss function. The batch size was set to 256, and the learning rate was adjusted using a learning rate scheduler, which started at 0.01, decreased to 0.001 after 3 epochs, and further reduced to 0.0001 after 30 epochs. Moreover, an early stopping mechanism was adopted to prevent overfitting. Specifically, the training was halted when the loss did not decrease for 15 consecutive training epochs, and the model parameters with the lowest loss were saved. The deep learning tasks were executed using Python (version 3.9) and the PyTorch library (version 2.1.0) on Linux (Ubuntu 22.04.4 LTS). A single Nvidia GeForce RTX 3090 GPU was used on the hardware front for training and inference. 

\subsection{Evaluation criteria}
This study assessed sEMG contaminant removal methods using two categories of evaluation metrics: signal reconstruction quality and feature extraction accuracy. The first group included SNR improvement (SNR$_{imp}$), root-mean-square error (RMSE), and percentage root-mean-square difference (PRD). These metrics have been broadly used to assess the outcomes of signal enhancement studies~\cite{chiang2019noise,zhang2021eegdenoisenet,sauer2024signal}. In the following equations, $x[n],\tilde x[n],\text{and }\hat x[n]$ denote clean, noisy, and enhanced sEMG waveform, respectively.

SNR$_{imp}$ represents the difference between the SNR after noise reduction and the original input signal SNR. Higher SNR$_{imp}$ values indicate better signal quality. The calculation is performed using the following equations:
\begin{equation}
\label{eq:SNRimp}
    \text{SNR}_{imp}=\text{SNR}_{out}-\text{SNR}_{in} ,
\end{equation}
where SNR$_{out}$ and SNR$_{in}$ denote the SNR values of the output and input sEMG signals, respectively, and are defined as

\begin{equation}
\label{eq:SNRout}
\text{SNR}_{out}=10\log_{10}{\bigg(\frac{\sum_{n=1}^{N}x[n]^2}{\sum_{n=1}^{N}(x[n]-\hat x[n])^2}\bigg)},
\end{equation}
\begin{equation}
\label{eq:SNRin}
    \text{SNR}_{in} = 10\log_{10}{\bigg(\frac{\sum_{n=1}^{N}x[n]^2}{\sum_{n=1}^{N}(x[n]-\tilde x[n])^2}\bigg)}.
\end{equation}

The RMSE indicates the average difference between the reconstructed output and the ground truth. Lower RMSE values indicate a better signal reconstruction quality. The RMSE is calculated using the following equation:
\begin{equation}
\label{eq:RMSE}
    \text{RMSE}=\sqrt{\frac{\sum_{n=1}^{N}(x[n]-\hat x[n])^2}{N}},
\end{equation}

The PRD calculates the percentage of the root-mean-square difference between the clean signal and enhanced signals, avoiding the influence of the clean signal energy level. Lower PRD values indicate a better signal reconstruction quality. The equation for PRD is
\begin{equation}
\label{eq:PRD}
\text{PRD} = \frac{\sqrt{\sum_{n=1}^{N}(x[n] - \hat{x}[n])^2}}{\sqrt{\sum_{n=1}^{N} x[n]^2}} \times 100.
\end{equation}

The second group of metrics, the feature extraction error, includes the RMSE of the average rectified value (ARV) and mean frequency (MF) feature vectors extracted from noisy and enhanced sEMG. Both the ARV and MF are often used in sEMG applications for clinical evaluation and diagnosis, including trunk muscle fatigue assessment~\cite{moniri2020real,phinyomark2012usefulness}, muscle force evaluation~\cite{phinyomark2012usefulness,abdelouahad2018time}, and muscle compensation detection in rehabilitation~\cite{ma2019semg}. The ARV is calculated as follows:
\begin{equation}
\label{eq:ARV}
    \text{ARV} = \frac{\sum_{n=1}^{L}|x[n]|}{L}.
\end{equation}
where $L$ is the sliding window length. In this study, $L$ was set to 200 to calculate the ARV with a 200-ms sliding window without overlap~\cite{liu2022research}. 

The MF characterizes the power spectrum distribution. This paper defines the MF as the expected value of the STFT amplitude spectrum between 10 and 500 Hz~\cite{xu2020comparative}. The calculation can be expressed as
\begin{equation}
\label{eq:MF}
    \text{MF} = \frac{\sum_{n=N_1}^{N_2}f_n\cdot S_n}{\sum_{n=N_1}^{N_2}S_n}.
\end{equation}
where $f_n$ and $S_n$ are the frequency and amplitude of the sEMG spectrogram, respectively. The sliding window in STFT also has a non-overlapping 200-ms window length, as in the ARV. Moreover, the MF is computed exclusively during the activation duration when exercises are performed~\cite{xu2020comparative}. 

For the ARV and MF, lower RMSE values of the extracted feature vectors indicate better performance robustness in applications using these sEMG features.

\begin{figure}[t!]
    \centering
    \includegraphics[width=.6\columnwidth]{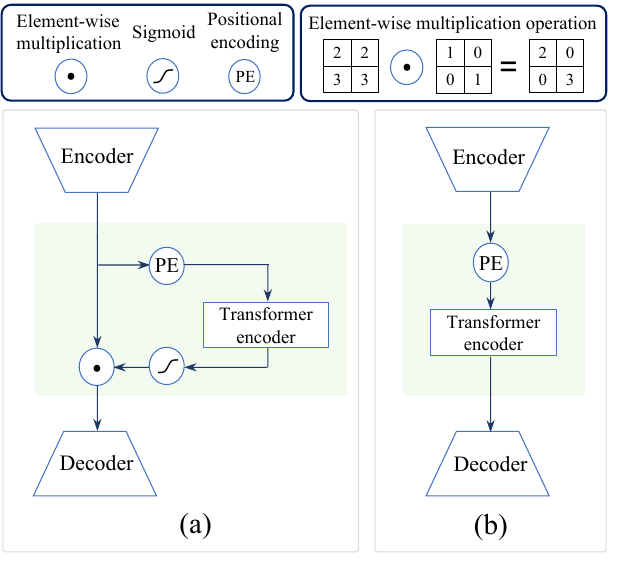}
    \caption{Illustration of the (a) representation masking and (b) direct mapping approaches.}
    \label{fig:RM/DM}
\end{figure}

\section{Results}
This study analyzed the experimental results from three perspectives: overall performance, performance under different SNR inputs, and performance with different types of contaminants. In the first subsection, the proposed method is compared with five existing sEMG denoising algorithms: IIR filters~\cite{drake2006elimination,ma2020emg,boyer2023reducing}, TS with IIR filters~\cite{junior2019template,marker2014effects,drake2006elimination}, and EMD-based~\cite{ma2020emg, naji2011application}, CEEMDAN-based~\cite{sun2023emg,sun2020surface,zhang2016improved}, and VMD-based sEMG denoising methods~\cite{ma2020emg,ashraf2023variational}. The second subsection describes the ablation study and compares the proposed method with several CNN-based architectures (CNN, FCN, and U-Net) because convolutional layers are suitable for processing waveform input data. Moreover, a direct mapping (DM) approach for the Transformer encoder was implemented and denoted as TrustEMG-Net(DM) for comparison. The difference between the DM and RM is shown in Fig.~\ref{fig:RM/DM}. A detailed implementation of the compared methods and NN models is available on the GitHub page\footnote{\url{https://github.com/eric-wang135/TrustEMG/tree/main/doc}}.

\renewcommand{\arraystretch}{1.4}
\begin{table*}[]
    \scriptsize
    \centering
    \caption{Overall performance of sEMG denoising methods.}
    \label{tab:nonNN_overall}
    \setlength{\tabcolsep}{5.5pt}
    \begin{tabular}{cccccccc}
    \toprule
    Metrics & Noisy & IIR & TS+IIR & EMD & CEEMDAN & VMD & TrustEMG-Net \\
    \cmidrule{2-8} 
    SNR$_{\scaleto{imp}{4pt}}$ & - &8.47 $\pm$ 0.08* &8.27 $\pm$ 0.08* &10.40 $\pm$ 0.09* &10.68 $\pm$ 0.09* &10.60 $\pm$ 0.15* &\bf 13.64 $\pm$ 0.38 \\
    RMSE ($\times 10^{\scaleto{-2}{3pt}}$) & 10.93 $\pm$ 1.10* &4.55 $\pm$ 0.49* &4.62 $\pm$ 0.50* & \phantom{*}3.05 $\pm$ 0.31* & \phantom{*}2.95 $\pm$ 0.30* & \phantom{*}2.95 $\pm$ 0.32* & \phantom{*}\bf 2.18 $\pm$ 0.28 \\
    PRD (\%) & 244.45 $\pm$ 0.00*\phantom{*} & 102.95 $\pm$ 0.60*\phantom{*} &104.36 $\pm$ 0.65* &69.30 $\pm$ 0.62* &67.14 $\pm$ 0.63* &67.00 $\pm$ 1.00* &\bf 48.44 $\pm$ 1.70 \\
    RMSE of ARV ($\times 10^{\scaleto{-3}{3pt}}$) & 69.75 $\pm$ 6.96* &20.33 $\pm$ 2.12* &20.43 $\pm$ 2.15* &12.51 $\pm$ 1.23* &11.55 $\pm$ 1.15* &11.08 $\pm$ 1.18* & \phantom{*}\bf 8.72 $\pm$ 1.52 \\
    RMSE of MF (Hz) & 42.25 $\pm$ 1.24* &59.30 $\pm$ 1.22* &61.15 $\pm$ 1.27* &36.05 $\pm$ 1.59* &32.14 $\pm$ 1.46* &38.26 $\pm$ 0.93* &\bf 16.63 $\pm$ 0.91 \\
    \bottomrule
    \multicolumn{8}{l}{\footnotesize *Denotes a significant difference (\textit{p}-value $<$ 0.05) with the proposed method. \textbf{Bold} font indicates the best score for each metric.}
    \end{tabular}
\end{table*}

\begin{figure*}
    \centering
    \includegraphics[width=\textwidth]{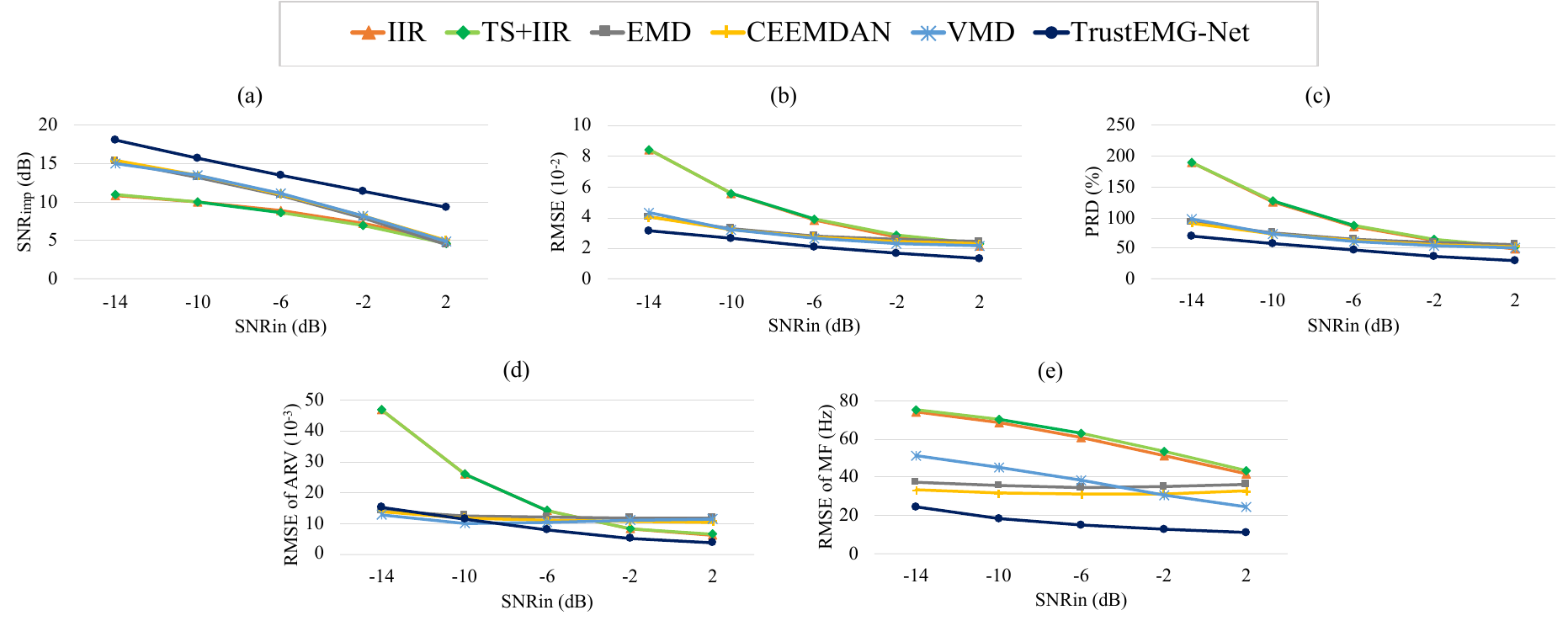}
    \caption{Performance under different SNR inputs measured using (a) SNR$_{imp}$, (b) RMSE, (c) PRD, (d) RMSE of the ARV, and (e) RMSE of the MF.}
    \label{fig:nonNN_different SNR}
\end{figure*}

\renewcommand{\arraystretch}{1.0}
\begin{table*}[]
    \scriptsize
    \centering
    \caption{Performance of sEMG denoising methods under different contaminant types.}
    \label{tab:nonNN_different noise}
    \begin{tabular}{ccccccccc}
    \toprule
    \multirow{2}{*}{Metric} & \multirow{2}{*}{Method} & \multicolumn{7}{c} {Contaminant type} \\
     \cmidrule{3-9}
     &  & BW & PLI & ECG & MOA & WGN & 3 types & 5 types \\
     \midrule 
    & IIR & 19.72 $\pm$ 0.09* & 15.73 $\pm$ 0.20* & 12.78 $\pm$ 0.28* & 14.06 $\pm$ 0.12* & 0.17 $\pm$ 0.00* & 7.45 $\pm$ 0.09* & 5.46 $\pm$ 0.07* \\ 
    & TS+IIR & 19.72 $\pm$ 0.09* & 15.73 $\pm$ 0.20* & 12.20 $\pm$ 0.29* & 14.06 $\pm$ 0.12* & 0.17 $\pm$ 0.00* & 7.16 $\pm$ 0.09* & 5.26 $\pm$ 0.08* \\ 
    SNR$_{imp} $& EMD & 20.25 $\pm$ 0.14* & 17.04 $\pm$ 0.08* & 12.58 $\pm$ 0.22* & 13.59 $\pm$ 0.22* & 6.42 $\pm$ 0.08* & 9.70 $\pm$ 0.10* & 7.51 $\pm$ 0.09* \\ 
    (dB) & CEEMDAN & \bf 20.61 $\pm$ 0.06* & 17.22 $\pm$ 0.09* & 12.55 $\pm$ 0.21* & 13.87 $\pm$ 0.13* & 6.51 $\pm$ 0.07* & 9.98 $\pm$ 0.11* & 7.90 $\pm$ 0.10* \\ 
    & VMD & 18.37 $\pm$ 0.27* & 17.35 $\pm$ 0.10* & 12.78 $\pm$ 0.28* & 13.44 $\pm$ 0.19* & 6.40 $\pm$ 0.06* & 9.86 $\pm$ 0.17* & 8.28 $\pm$ 0.16* \\ 
    & TrustEMG-Net & 18.94 $\pm$ 0.90\phantom{*} & \bf 21.45 $\pm$ 0.70\phantom{*} & \bf 17.19 $\pm$ 0.77\phantom{*} & \bf 15.89 $\pm$ 0.71\phantom{*} & \bf 9.06 $\pm$ 0.32\phantom{*} & \bf 13.09 $\pm$ 0.35\phantom{*} & \bf 11.38 $\pm$ 0.22\phantom{*} \\
    \midrule
     & IIR & 1.21 $\pm$ 0.11* & 1.50 $\pm$ 0.18* & 2.08 $\pm$ 0.24* & 2.37 $\pm$ 0.26* & 10.64 $\pm$ 1.13* & 4.79 $\pm$ 0.50* & 5.32 $\pm$ 0.57* \\ 
    & TS+IIR & 1.21 $\pm$ 0.11* & 1.50 $\pm$ 0.18* & 2.19 $\pm$ 0.26* & 2.37 $\pm$ 0.26* & 10.64 $\pm$ 1.13* & 4.88 $\pm$ 0.51* & 5.40 $\pm$ 0.58* \\ 
    RMSE & EMD & 1.22 $\pm$ 0.11* & 1.40 $\pm$ 0.15* & 2.23 $\pm$ 0.24* & 2.45 $\pm$ 0.26* & 4.38 $\pm$ 0.44* & 3.11 $\pm$ 0.31* & 3.72 $\pm$ 0.37* \\ 
    ($\times 10^{\scaleto{-2}{3pt}}$) & CEEMDAN & \bf 1.14 $\pm$ 0.11* & 1.36 $\pm$ 0.15* & 2.24 $\pm$ 0.24* & 2.39 $\pm$ 0.26* & 4.38 $\pm$ 0.44* & 3.00 $\pm$ 0.30* & 3.55 $\pm$ 0.36* \\ 
    & VMD & 1.17 $\pm$ 0.10* & 1.29 $\pm$ 0.14* & 2.08 $\pm$ 0.24* & 2.34 $\pm$ 0.26* & 4.66 $\pm$ 0.49* & 3.06 $\pm$ 0.33* & 3.49 $\pm$ 0.38* \\ 
    & TrustEMG-Net & 1.31 $\pm$ 0.26\phantom{*} & \bf 0.85 $\pm$ 0.18\phantom{*} & \bf 1.45 $\pm$ 0.27\phantom{*} & \bf 1.77 $\pm$ 0.28\phantom{*} & \bf 3.33 $\pm$ 0.39\phantom{*} & \bf 2.26 $\pm$ 0.28\phantom{*} & \bf 2.56 $\pm$ 0.30\phantom{*} \\ 
    \midrule
     & IIR & 27.59 $\pm$ 0.29\phantom{*} & 33.40 $\pm$ 0.77* & 47.90 $\pm$ 1.41* & 53.88 $\pm$ 0.65* & 239.49 $\pm$ 1.80* & 108.27 $\pm$ 0.69* & 120.27 $\pm$ 0.72* \\ 
    & TS+IIR & 27.59 $\pm$ 0.29\phantom{*} & 33.40 $\pm$ 0.77* & 50.42 $\pm$ 1.59* & 53.88 $\pm$ 0.65* & 239.49 $\pm$ 1.80* & 110.17 $\pm$ 0.71* & 122.08 $\pm$ 0.76* \\ 
    PRD & EMD & 27.08 $\pm$ 0.62\phantom{*} & 31.64 $\pm$ 0.32* & 51.03 $\pm$ 1.07* & 55.31 $\pm$ 0.99* & 98.68 $\pm$ 0.74* & 70.61 $\pm$ 0.67* & 84.56 $\pm$ 0.82* \\ 
    (\%) & CEEMDAN & \bf 25.82 $\pm$ 0.26* & 30.45 $\pm$ 0.29* & 51.22 $\pm$ 1.04* & 54.46 $\pm$ 0.71* & 98.74 $\pm$ 0.76* & 68.25 $\pm$ 0.72* & 81.07 $\pm$ 0.90* \\ 
    & VMD & 28.10 $\pm$ 0.79\phantom{*} & 29.19 $\pm$ 0.36* & 47.90 $\pm$ 1.41* & 53.81 $\pm$ 0.81* & 104.38 $\pm$ 0.88* & 69.53 $\pm$ 1.13* & 78.81 $\pm$ 1.25* \\ 
    & TrustEMG-Net & 27.66 $\pm$ 3.10\phantom{*} & \bf 18.33 $\pm$ 1.80\phantom{*} & \bf 30.96 $\pm$ 2.93\phantom{*} & \bf 38.35 $\pm$ 2.44\phantom{*} & \bf 74.32 $\pm$ 1.95\phantom{*} & \bf 50.25 $\pm$ 1.58\phantom{*} & \bf 57.42 $\pm$ 1.23\phantom{*} \\
    \midrule
    & IIR & 3.52 $\pm$ 0.34* & 3.89 $\pm$ 0.51* & 5.74 $\pm$ 0.75\phantom{*} & 7.27 $\pm$ 0.80* & 65.02 $\pm$ 6.79* & 22.03 $\pm$ 2.24* & 21.89 $\pm$ 2.25* \\ 
    & TS+IIR & 3.52 $\pm$ 0.34* & 3.89 $\pm$ 0.51* & 6.44 $\pm$ 0.93* & 7.27 $\pm$ 0.80* & 65.02 $\pm$ 6.79* & 22.22 $\pm$ 2.28* & 21.88 $\pm$ 2.26* \\ 
    RMSE of ARV & EMD & 3.48 $\pm$ 0.32* & 4.58 $\pm$ 0.47* & 6.21 $\pm$ 0.68* & 7.06 $\pm$ 0.77* & 13.96 $\pm$ 1.31* & 12.40 $\pm$ 1.23* & 18.09 $\pm$ 1.77* \\ 
    ($\times 10^{\scaleto{-3}{3pt}}$) & CEEMDAN & \bf 3.36 $\pm$ 0.33* & 4.36 $\pm$ 0.45* & 6.25 $\pm$ 0.68* & 6.91 $\pm$ 0.78\phantom{*} & \bf 13.44 $\pm$ 1.28* & 11.25 $\pm$ 1.12* & 16.53 $\pm$ 1.63* \\ 
    & VMD & 3.59 $\pm$ 0.28* & 4.82 $\pm$ 0.49* & 5.74 $\pm$ 0.75\phantom{*} & 7.34 $\pm$ 0.77* & 16.39 $\pm$ 1.65\phantom{*} & 11.07 $\pm$ 1.18* & 14.60 $\pm$ 1.59* \\ 
    & TrustEMG-Net & 5.34 $\pm$ 1.77\phantom{*} & \bf 2.42 $\pm$ 1.21\phantom{*} & \bf 5.41 $\pm$ 1.59\phantom{*} & \bf 6.49 $\pm$ 1.57\phantom{*} & 16.85 $\pm$ 2.31\phantom{*} & \bf 9.08 $\pm$ 1.53\phantom{*} & \bf 9.84 $\pm$ 1.49\phantom{*} \\
    \midrule
     & IIR & 7.55 $\pm$ 1.18* & 8.59 $\pm$ 0.27* & 15.97 $\pm$ 0.45* & \bf 12.99 $\pm$ 0.74\phantom{*} & 90.89 $\pm$ 2.32* & 60.60 $\pm$ 1.20* & 90.18 $\pm$ 2.07* \\ 
    & TS+IIR & 7.55 $\pm$ 1.18* & 8.59 $\pm$ 0.27* & 20.27 $\pm$ 1.14* & \bf 12.99 $\pm$ 0.74\phantom{*} & 90.89 $\pm$ 2.32* & 62.93 $\pm$ 1.27* & 92.54 $\pm$ 2.12* \\ 
    RMSE of MF & EMD & 8.80 $\pm$ 1.03* & 6.08 $\pm$ 0.43* & 14.27 $\pm$ 0.64* & 15.34 $\pm$ 0.60* & 58.72 $\pm$ 2.56* & 37.57 $\pm$ 1.58* & 49.99 $\pm$ 2.42* \\ 
    (Hz) & CEEMDAN & \bf 7.41 $\pm$ 1.15* & 5.93 $\pm$ 0.42* & 14.33 $\pm$ 0.65* & 13.79 $\pm$ 0.58\phantom{*} & 53.05 $\pm$ 2.43* & 33.57 $\pm$ 1.45* & 43.99 $\pm$ 2.23* \\ 
    & VMD & 8.12 $\pm$ 0.77* & 6.44 $\pm$ 0.47* & 15.97 $\pm$ 0.45* & 13.39 $\pm$ 0.73\phantom{*} & 58.09 $\pm$ 2.01* & 40.71 $\pm$ 0.91* & 53.69 $\pm$ 1.58* \\ 
    & TrustEMG-Net & 10.98 $\pm$ 2.92\phantom{*} & \bf 4.24 $\pm$ 1.40\phantom{*} & \bf 11.67 $\pm$ 2.47\phantom{*} & 14.04 $\pm$ 2.55\phantom{*} & \bf 29.11 $\pm$ 2.65\phantom{*} & \bf 16.92 $\pm$ 0.92\phantom{*} & \bf 19.07 $\pm$ 0.99\phantom{*} \\
    \bottomrule
    \multicolumn{9}{l}{\footnotesize *Denotes a significant difference (\textit{p}-value $<$ 0.05) with the proposed method. \textbf{Bold} font indicates the best score for each metric.}
    \end{tabular}
\end{table*}

\renewcommand{\arraystretch}{1.4}
\begin{table*}[]
    \centering
    \caption{Overall performance of NN models for sEMG contaminant removal.}
    \label{tab:NN_overall}
    \begin{tabular}{ccccccc}
    \toprule
    Metric & Noisy & CNN & FCN & U-Net & TrustEMG-Net(DM) & TrustEMG-Net \\
    \cmidrule{2-7}
    SNR$_{imp}$ (dB) & - & \phantom{*}8.98 $\pm$ 0.72* &11.56 $\pm$ 0.26* &13.16 $\pm$ 0.44* &13.45 $\pm$ 0.45* &\bf 13.64 $\pm$ 0.38 \\
    RMSE ($\times 10^{\scaleto{-2}{3pt}}$)  & 10.93 $\pm$ 1.10* & \phantom{*}3.22 $\pm$ 0.39* & \phantom{*}2.51 $\pm$ 0.29* & \phantom{*}2.27 $\pm$ 0.31* & \phantom{*}2.21 $\pm$ 0.30* & \phantom{1}\bf 2.18 $\pm$ 0.28 \\
    PRD (\%) & 244.45 $\pm$ 0.00*\phantom{2} &72.70 $\pm$ 6.37* &56.97 $\pm$ 1.58* &50.19 $\pm$ 1.95* &49.05 $\pm$ 2.00* &\bf 48.44 $\pm$ 1.70 \\
    RMSE of ARV ($\times 10^{\scaleto{-3}{3pt}}$)   & 69.75 $\pm$ 6.96* &18.32 $\pm$ 2.83* &11.32 $\pm$ 1.68* & \phantom{*}9.36 $\pm$ 1.73* & \phantom{*}8.90 $\pm$ 1.58* & \phantom{1}\bf 8.72 $\pm$ 1.52 \\
    RMSE of MF (Hz) & 42.25 $\pm$ 1.24* &23.80 $\pm$ 6.25* & \bf 16.24 $\pm$ 0.76* & 17.62 $\pm$ 1.04* &16.94 $\pm$ 0.94* & 16.63 $\pm$ 0.91 \\
    \bottomrule
    \multicolumn{7}{l}{\footnotesize *Denotes a significant difference (\textit{p}-value $<$ 0.05) with the proposed method. \textbf{Bold} font indicates the best score for each metric.}
    \end{tabular}
\end{table*}

\renewcommand{\arraystretch}{.9}
\begin{table*}[]
    \centering
    \scriptsize
    \caption{Performance of NN models under different SNR inputs.}
    \label{tab:NN_different SNR}
    \begin{tabularx}{\textwidth}{*{7}Y}
    \toprule
    \multirow{2}{*}{Metric} & \multirow{2}{*}{Model} & \multicolumn{5}{c} {SNR level} \\
    \cmidrule{3-7}
     & & -14 dB & -10 dB & -6 dB & -2 dB & 2 dB \\
    \midrule
    & CNN & 15.69 $\pm$ 0.42* & 12.32 $\pm$ 0.56* & \phantom{1}9.03 $\pm$ 0.73* &  \phantom{1}5.68 $\pm$ 0.90* & 2.16 $\pm$ 1.02* \\ 
    & FCN & 17.11 $\pm$ 0.20* & 14.28 $\pm$ 0.23* & 11.57 $\pm$ 0.27* & \phantom{1}8.86 $\pm$ 0.31* & 6.00 $\pm$ 0.33* \\ 
    SNR$_{imp} $& U-Net & 17.96 $\pm$ 0.27* & 15.42 $\pm$ 0.36* & 13.11 $\pm$ 0.44* & 10.88 $\pm$ 0.53* & 8.42 $\pm$ 0.60* \\ 
    (dB) & TrustEMG-Net(DM) & 18.05 $\pm$ 0.26* & 15.59 $\pm$ 0.35* & 13.38 $\pm$ 0.44* & 11.26 $\pm$ 0.56* & 8.99 $\pm$ 0.66* \\ 
    & TrustEMG-Net & \bf 18.13 $\pm$ 0.26\phantom{*} & \bf 15.69 $\pm$ 0.34\phantom{*} & \bf 13.53 $\pm$ 0.41\phantom{*} & \bf 11.51 $\pm$ 0.46\phantom{*} & \bf 9.36 $\pm$ 0.46\phantom{*} \\
    \midrule
    & CNN & 3.73 $\pm$ 0.40* & 3.48 $\pm$ 0.40* & 3.19 $\pm$ 0.39* & 2.93 $\pm$ 0.40* & 2.77 $\pm$ 0.40* \\ 
    & FCN & 3.32 $\pm$ 0.37* & 2.89 $\pm$ 0.33* & 2.46 $\pm$ 0.29* & 2.08 $\pm$ 0.25* & 1.81 $\pm$ 0.23* \\ 
    RMSE & U-Net & 3.16 $\pm$ 0.37* & 2.70 $\pm$ 0.34* & 2.21 $\pm$ 0.30* & 1.79 $\pm$ 0.27* & 1.50 $\pm$ 0.26* \\ 
    ($\times 10^{\scaleto{-2}{3pt}}$) & TrustEMG-Net(DM) & 3.13 $\pm$ 0.37* & 2.66 $\pm$ 0.33* & 2.16 $\pm$ 0.29* & 1.72 $\pm$ 0.26* & 1.40 $\pm$ 0.24* \\ 
    & TrustEMG-Net & \bf 3.12 $\pm$ 0.36\phantom{*} & \bf 2.64 $\pm$ 0.33\phantom{*} & \bf 2.13 $\pm$ 0.28\phantom{*} & \bf 1.68 $\pm$ 0.24\phantom{*} & \bf 1.34 $\pm$ 0.20\phantom{*} \\
    \midrule
    & CNN & 84.19 $\pm$ 3.90* & 78.16 $\pm$ 5.18* & 71.83 $\pm$ 6.57* & 66.44 $\pm$ 7.79* & 62.90 $\pm$ 8.61* \\ 
    & FCN & 75.07 $\pm$ 1.94* & 65.39 $\pm$ 1.72* & 55.76 $\pm$ 1.61* & 47.40 $\pm$ 1.53* & 41.21 $\pm$ 1.42* \\ 
    PRD & U-Net & 70.28 $\pm$ 1.50* & 59.67 $\pm$ 1.87* & 48.84 $\pm$ 2.05* & 39.46 $\pm$ 2.17* & 32.71 $\pm$ 2.23* \\ 
    (\%) & TrustEMG-Net(DM) & 69.57 $\pm$ 1.51* & 58.77 $\pm$ 1.87* & 47.74 $\pm$ 2.09* & 38.18 $\pm$ 2.26* & 30.99 $\pm$ 2.36\phantom{*} \\
    & TrustEMG-Net & \bf 69.36 $\pm$ 1.49\phantom{*} & \bf 58.47 $\pm$ 1.78\phantom{*} & \bf 47.25 $\pm$ 1.89\phantom{*} & \bf 37.33 $\pm$ 1.83\phantom{*} & \bf 29.80 $\pm$ 1.55\phantom{*} \\
    \midrule
    & CNN & 22.49 $\pm$ 2.90* & 20.32 $\pm$ 2.88* & 17.80 $\pm$ 2.92* & 15.92 $\pm$ 2.99*\phantom{1} & 15.05 $\pm$ 3.04*\phantom{1} \\ 
    & FCN & 16.57 $\pm$ 2.10* & 13.79 $\pm$ 1.98* & 10.66 $\pm$ 1.68* & 8.41 $\pm$ 1.42* & 7.19 $\pm$ 1.26* \\ 
    RMSE of ARV & U-Net & 15.25 $\pm$ 2.16* & 11.88 $\pm$ 1.97* & \phantom{1}8.49 $\pm$ 1.67* & 6.20 $\pm$ 1.52* & 5.00 $\pm$ 1.46\phantom{*} \\
    ($\times 10^{\scaleto{-3}{3pt}}$) & TrustEMG-Net(DM) & \bf 14.97 $\pm$ 1.92* & 11.55 $\pm$ 1.79* & \phantom{1}8.00 $\pm$ 1.57* & 5.64 $\pm$ 1.44\phantom{*} & 4.34 $\pm$ 1.37* \\ 
    & TrustEMG-Net & 15.23 $\pm$ 1.99\phantom{*} & \bf 11.50 $\pm$ 1.84\phantom{*} & \phantom{1}\bf 7.78 $\pm$ 1.57\phantom{*} & \bf 5.26 $\pm$ 1.30\phantom{*} & \bf 3.80 $\pm$ 1.03\phantom{*} \\
    \midrule
    & CNN & 27.40 $\pm$ 5.82* & 24.85 $\pm$ 6.08* & 23.05 $\pm$ 6.46* & 22.05 $\pm$ 6.73* & 21.66 $\pm$ 6.84\phantom{*} \\
    & FCN & \bf 22.48 $\pm$ 1.31* & \bf 17.89 $\pm$ 0.91* & 15.17 $\pm$ 0.80\phantom{*} & 13.41 $\pm$ 0.69* & 12.25 $\pm$ 0.53* \\ 
    RMSE of MF & U-Net & 23.80 $\pm$ 1.47* & 19.05 $\pm$ 1.02* & 16.43 $\pm$ 0.95* & 14.85 $\pm$ 1.04* & 13.96 $\pm$ 1.32* \\ 
    (Hz) & TrustEMG-Net(DM) & 24.38 $\pm$ 1.70* & 18.84 $\pm$ 0.99* & 15.51 $\pm$ 0.77* & 13.60 $\pm$ 0.94\phantom{*} & 12.38 $\pm$ 1.32\phantom{*} \\
    & TrustEMG-Net & 24.76 $\pm$ 1.90\phantom{*} & 18.64 $\pm$ 1.05\phantom{*} & \bf 15.16 $\pm$ 0.85\phantom{*} & \bf 13.06 $\pm$ 0.81\phantom{*} & \bf 11.54 $\pm$ 0.78\phantom{*} \\
    \bottomrule
    \multicolumn{7}{l}{\footnotesize *Denotes a significant difference (\textit{p}-value $<$ 0.05) with the proposed method. \textbf{Bold} font indicates the best score for each metric.}\\
    \end{tabularx}
\end{table*}

\begin{table*}[]
    \centering
    \scriptsize
    \caption{Performance of NN models under different contaminant types.}
    \label{tab:NN_different noise}
    \begin{tabular}{ccccccccc}
    \toprule
    \multirow{2}{*}{Metric} & \multirow{2}{*}{Model} & \multicolumn{7}{c} {Contaminant type} \\
    \cmidrule{3-9}
     &  & BW & PLI & ECG & MOA & WGN & 3 types & 5 types \\
    \midrule
    & CNN & 10.17 $\pm$ 1.05* & 10.48 $\pm$ 1.12* & 10.11 $\pm$ 1.05* & \phantom{1}9.85 $\pm$ 1.09* & 7.40 $\pm$ 0.53* & \phantom{1}8.89 $\pm$ 0.72* & \phantom{1}8.52 $\pm$ 0.60* \\ 
    & FCN & 14.73 $\pm$ 0.37* & 15.84 $\pm$ 0.47* & 14.14 $\pm$ 0.34* & 13.43 $\pm$ 0.53* & 8.26 $\pm$ 0.39* & 11.29 $\pm$ 0.27* & 10.18 $\pm$ 0.24* \\ 
    SNR$_{imp} $& U-Net & 17.65 $\pm$ 1.05* & 20.86 $\pm$ 0.73* & 16.06 $\pm$ 0.82* & 14.73 $\pm$ 0.80* & 9.01 $\pm$ 0.33* & 12.64 $\pm$ 0.39* & 11.22 $\pm$ 0.26* \\ 
    (dB) & TrustEMG-Net(DM) & 18.48 $\pm$ 1.06* & 20.90 $\pm$ 0.72* & 16.96 $\pm$ 0.81* & 15.59 $\pm$ 0.84* & 9.05 $\pm$ 0.33\phantom{*} & 12.91 $\pm$ 0.42* & 11.29 $\pm$ 0.26* \\ 
    & TrustEMG-Net & \bf 18.94 $\pm$ 0.90\phantom{*} & \bf 21.45 $\pm$ 0.70\phantom{*} & \bf 17.19 $\pm$ 0.77\phantom{*} & \bf 15.89 $\pm$ 0.71\phantom{*} & \bf 9.06 $\pm$ 0.32\phantom{*} & \bf 13.09 $\pm$ 0.35\phantom{*} & \bf 11.38 $\pm$ 0.22\phantom{*} \\
    \midrule
    & CNN & 2.78 $\pm$ 0.41* & 2.68 $\pm$ 0.41* & 2.81 $\pm$ 0.41* & 2.95 $\pm$ 0.41* & 3.93 $\pm$ 0.45* & 3.26 $\pm$ 0.39* & 3.37 $\pm$ 0.39* \\ 
    & FCN & 1.72 $\pm$ 0.22* & 1.47 $\pm$ 0.20* & 1.80 $\pm$ 0.24* & 2.03 $\pm$ 0.26* & 3.62 $\pm$ 0.41* & 2.57 $\pm$ 0.30* & 2.84 $\pm$ 0.32* \\ 
    RMSE & U-Net & 1.49 $\pm$ 0.33* & 0.92 $\pm$ 0.20* & 1.66 $\pm$ 0.32* & 1.99 $\pm$ 0.34* & 3.34 $\pm$ 0.38* & 2.35 $\pm$ 0.31* & 2.60 $\pm$ 0.31* \\ 
    ($\times 10^{\scaleto{-2}{3pt}}$) & TrustEMG-Net(DM) & 1.39 $\pm$ 0.33* & 0.90 $\pm$ 0.18* & 1.48 $\pm$ 0.29* & 1.82 $\pm$ 0.33* & 3.33 $\pm$ 0.39\phantom{*} & 2.29 $\pm$ 0.30* & 2.58 $\pm$ 0.30* \\ 
    & TrustEMG-Net & \bf 1.31 $\pm$ 0.26\phantom{*} & \bf 0.85 $\pm$ 0.18\phantom{*} & \bf 1.45 $\pm$ 0.27\phantom{*} & \bf 1.77 $\pm$ 0.28\phantom{*} & \bf 3.33 $\pm$ 0.39\phantom{*} & \bf 2.26 $\pm$ 0.28\phantom{*} & \bf 2.56 $\pm$ 0.30\phantom{*} \\
    \midrule
    & CNN & 62.86 $\pm$ 8.75* & 61.00 $\pm$ 8.97* & 63.41 $\pm$ 8.66* & 66.71 $\pm$ 7.80* & 87.67 $\pm$ 3.49* & 73.65 $\pm$ 6.15* & 76.28 $\pm$ 5.54* \\ 
    & FCN & 38.81 $\pm$ 1.37* & 33.56 $\pm$ 1.50* & 40.83 $\pm$ 1.52* & 46.34 $\pm$ 1.37* & 80.78 $\pm$ 2.21* & 58.48 $\pm$ 1.58* & 64.58 $\pm$ 1.68* \\ 
    PRD & U-Net & 31.10 $\pm$ 3.77* & 19.60 $\pm$ 1.94* & 34.87 $\pm$ 3.35* & 42.66 $\pm$ 2.88* & 74.61 $\pm$ 1.95* & 51.99 $\pm$ 1.87* & 58.27 $\pm$ 1.47* \\ 
    (\%) & TrustEMG-Net(DM) & 29.00 $\pm$ 3.78* & 19.42 $\pm$ 1.88* & 31.68 $\pm$ 3.18* & 39.22 $\pm$ 3.16* & \bf 74.29 $\pm$ 2.04\phantom{*} & 50.87 $\pm$ 1.94* & 57.82 $\pm$ 1.47* \\ 
    & TrustEMG-Net & \bf 27.66 $\pm$ 3.10\phantom{*} & \bf 18.33 $\pm$ 1.80\phantom{*} & \bf 30.96 $\pm$ 2.93\phantom{*} & \bf 38.35 $\pm$ 2.44\phantom{*} & 74.32 $\pm$ 1.95\phantom{*} & \bf 50.25 $\pm$ 1.58\phantom{*} & \bf 57.42 $\pm$ 1.23\phantom{*} \\
    \midrule
     & CNN & 15.39 $\pm$ 3.11*\phantom{1} & 14.31 $\pm$ 2.99*\phantom{1} & 15.88 $\pm$ 3.08* & 16.14 $\pm$ 3.04*\phantom{1} & 25.22 $\pm$ 3.45* & 18.63 $\pm$ 2.83* & 18.99 $\pm$ 2.83* \\ 
    & FCN & 7.03 $\pm$ 1.22* & 6.08 $\pm$ 1.09* & 7.66 $\pm$ 1.41* & 8.06 $\pm$ 1.46* & 18.93 $\pm$ 2.44* & 11.68 $\pm$ 1.70* & 12.80 $\pm$ 1.84* \\ 
    RMSE of ARV & U-Net & 6.46 $\pm$ 2.16* & 2.83 $\pm$ 1.28* & 7.21 $\pm$ 1.97* & 8.27 $\pm$ 2.03* & 16.76 $\pm$ 2.28\phantom{*} & \phantom{1}9.68 $\pm$ 1.74* & 10.17 $\pm$ 1.60* \\ 
    ($\times 10^{\scaleto{-3}{3pt}}$) & TrustEMG-Net(DM) & 5.69 $\pm$ 2.18\phantom{*} & 2.68 $\pm$ 1.14* & 5.74 $\pm$ 1.73* & 6.99 $\pm$ 1.97* & \bf 16.39 $\pm$ 2.21* &\phantom{1}9.23 $\pm$ 1.57\phantom{*} & 10.05 $\pm$ 1.46* \\ 
    & TrustEMG-Net & \bf 5.34 $\pm$ 1.77\phantom{*} & \bf 2.42 $\pm$ 1.21\phantom{*} & \bf 5.41 $\pm$ 1.59\phantom{*} & \bf 6.49 $\pm$ 1.57\phantom{*} & 16.85 $\pm$ 2.31\phantom{*} & \phantom{1}\bf 9.08 $\pm$ 1.53\phantom{*} & \phantom{1}\bf 9.84 $\pm$ 1.49\phantom{*} \\
    \midrule
    & CNN & 21.54 $\pm$ 6.93* & 20.17 $\pm$ 7.54*\phantom{1} & 20.71 $\pm$ 7.09* & 21.26 $\pm$ 7.00* & 32.87 $\pm$ 5.91* & 24.07 $\pm$ 6.22* & 24.09 $\pm$ 6.31* \\ 
    & FCN & \phantom{1}\bf 9.75 $\pm$ 0.41\phantom{*} & 6.64 $\pm$ 0.68* & \bf 10.73 $\pm$ 0.86* & \bf 12.09 $\pm$ 0.80* & 27.99 $\pm$ 1.95* & \bf 16.66 $\pm$ 0.82\phantom{*} & \bf 18.71 $\pm$ 1.04\phantom{*} \\
    RMSE of MF & U-Net & 12.17 $\pm$ 3.41* & 4.57 $\pm$ 1.46* & 14.50 $\pm$ 2.95* & 17.15 $\pm$ 2.94* & \bf 26.84 $\pm$ 2.18* & 17.92 $\pm$ 1.07* & 19.97 $\pm$ 1.00* \\ 
    (Hz) & TrustEMG-Net(DM) & 11.44 $\pm$ 3.26\phantom{*} & 4.69 $\pm$ 1.31* & 12.18 $\pm$ 2.68* & 15.01 $\pm$ 2.84* & 28.24 $\pm$ 2.27\phantom{*} & 17.28 $\pm$ 0.97* & 19.34 $\pm$ 1.31* \\ 
    & TrustEMG-Net & 10.98 $\pm$ 2.92\phantom{*} & \bf 4.24 $\pm$ 1.40\phantom{*} & 11.67 $\pm$ 2.47\phantom{*} & 14.04 $\pm$ 2.55\phantom{*} & 29.11 $\pm$ 2.65\phantom{*} & 16.92 $\pm$ 0.92\phantom{*} & 19.07 $\pm$ 0.99\phantom{*} \\
    \bottomrule
    \multicolumn{9}{l}{{\footnotesize *Denotes a significant difference (\textit{p}-value $<$ 0.05) with the proposed method. \textbf{Bold} font indicates the best score for each metric.}}
    \end{tabular}
\end{table*}

\subsection{Comparison with existing sEMG denoising methods}
\subsubsection{Overall performance}
Table~\ref{tab:nonNN_overall} presents the overall performance of the proposed and existing sEMG denoising methods. The proposed method achieved the highest score in all five metrics: an SNR$_{imp}$ of 13.64 dB, an RMSE of 2.18$\times 10^{-2}$, a PRD of 48.44\%, an RMSE of the ARV of 8.72$\times 10^{-3}$, and an RMSE of the MF of 16.63 Hz. Among the five existing denoising methods, the CEEMDAN- and VMD-based methods were comparable. The CEEMDAN-based method outperformed the others in terms of the SNR$_{imp}$, RMSE, and RMSE of the MF, whereas the VMD-based method yielded better results in terms of the PRD and RMSE of the ARV.

\subsubsection{Performance under different SNRs}
Fig.~\ref{fig:nonNN_different SNR} presents the performance metrics of the sEMG denoising methods under different SNRs. The proposed TrustEMG-Net, denoted by the dark blue line in the plots, achieved the best results for almost all the metrics and SNRs. The exception is the RMSE of the ARV, where the VMD-based method performs best at SNR -10 dB, and all decomposition-based methods perform slightly better than TrustEMG-Net at SNR -14 dB. Moreover, the performance differences between TrustEMG-Net and the other denoising methods, in most cases, were significantly different (\textit{p} $<0.05$). The only case without a significant difference wes between the CEEMDAN-based method and TrustEMG-Net in the RMSE of the ARV at SNR -10 dB (\textit{p} $=0.162$).

\subsubsection{Performance under different contaminant types}
Table~\ref{tab:nonNN_different noise} lists the sEMG denoising performance of the methods with different contaminant types. TrustEMG-Net significantly outperformed the other methods under most conditions. Exceptions included all metrics under BW and the RMSE of the ARV under WGN. The CEEMDAN-based method performs best under BW. Regarding the RMSE of the ARV under WGN, the four sEMG denoising methods outperformed TrustEMG-Net, but the results were not significantly different (\textit{p} $<0.05$).

\subsection{Ablation study}
\subsubsection{Overall performance}
Table~\ref{tab:NN_overall} shows the overall performance of the NN-based contaminant removal methods measured using five metrics. The proposed TrustEMG-Net consistently outperformed the other four NN models in four out of the five metrics (SNR$_{imp}$, RMSE, PRD, and RMSE of the ARV) with significant differences. For the RMSE of the MF, FCN performed slightly better than TrustEMG-Net. 

\subsubsection{Performance under different SNR inputs}
Table~\ref{tab:NN_different SNR} presents the performance of the NN-based sEMG denoising methods across different SNR levels. For three signal quality metrics, namely SNR$_{imp}$, RMSE, and PRD, TrustEMG-Net consistently outperformed other NN models across all SNR levels. Regarding the feature extraction error, TrustEMG-Net consistently yielded the lowest RMSE of the ARV across most SNRs. At an SNR of -14 dB, TrustEMG-Net(DM) achieved the lowest RMSE of the ARV of 14.97$\times 10^{-3}$, whereas TrustEMG-Net followed closely with values of 15.23$\times 10^{-3}$. For the RMSE of the MF, TrustEMG-Net produced the lowest error values at SNRs from -6 to 2 dB. However, FCN outperformed TrustEMG-Net at lower SNRs (-10 and -14 dB).

\subsubsection{Performance under different contaminant types}
Table~\ref{tab:NN_different noise} shows the performance of the NN-based denoising methods across various contaminant types. For the three signal quality metrics (SNR$_{imp}$, RMSE, and PRD), TrustEMG-Net significantly outperformed the other methods for all types of contaminants, except for WGN, where TrustEMG-Net(DM) exhibited comparable performance. As for the RMSE of the ARV, TrustEMG-Net was only significantly outperformed by TrustEMG-Net(DM) under WGN. Regarding the RMSE of the MF, the proposed approach performed best only under PLI. FCN outperformed TrustEMG-Net for six contaminant types, but the performance difference was only significant for three of them (ECG, WGN, and MOA). Additionally, U-Net yielded the lowest error under WGN for the RMSE of the MF.

\section{Discussion}
\label{sec:discussion}
\subsection{Comparison with existing sEMG denoising methods}
Our comparative evaluation with existing sEMG denoising algorithms underscores the effective denoising capability, robustness, and generalizability of TrustEMG-Net. 
Table~\ref{tab:nonNN_overall} demonstrates its significant overall outperformance, reaching up to a 20\%\textendash 48\% improvement across various metrics. Regarding the SNR inputs, Fig.~\ref{fig:nonNN_different SNR} shows the superiority of TrustEMG-Net under various SNR inputs, with its advantage being more pronounced under higher SNRs. The outperformance of the metrics can reach 38\%\textendash 83\% at an SNR of 2 dB. For various contaminant types, Table~\ref{tab:nonNN_different noise} indicates that the proposed approach excels, particularly under PLI and compound contaminant types, exhibiting improvements of 23\%\textendash 37\%, 17\%\textendash 49\%, and 26\%\textendash 56\% under PLI, three-type, and five-type contaminants, respectively. These enhancements validate the efficacy and novelty of TrustEMG-Net and demonstrate its robustness and superior denoising capabilities in a comprehensive range of scenarios.

The exceptional performance of the proposed method is attributed to its effective nonlinear mapping capability and completely data-driven nature. Nonlinear mapping enables TrustEMG-Net to effectively model the relationship between sEMG and contaminants in both the spectral and temporal domains effectively. In contrast, existing methods that rely solely on spectral or temporal features have limitations in handling diverse contaminants. For example, IIR filters in the frequency domain struggle with broadband contaminants, resulting in poor results when the sEMG is contaminated with WGN~\cite{sauer2024signal}. Similarly, methods such as TS, which function only in the temporal dimension, are effective for removing quasi-periodic contaminants but may falter in more complex scenarios. The nonlinear mapping mechanism of NN-based methods offers a more versatile approach resulting in an effective denoising capability and generalizability.

The completely data-driven nature of TrustEMG-Net further contributes to its superiority over existing methods. The feature extraction and contaminant removal processes are automatically optimized to minimize the signal reconstruction loss through gradient descent algorithms. This technique can achieve an optimal denoising performance with a representative training set encompassing various contaminant types and SNR levels, demonstrating robustness against SNR variations and generalizability to different contaminant types. In contrast, decomposition-based sEMG denoising methods often require human expertise and trial and error, making optimization challenging and potentially resulting in suboptimal results in more generalized scenarios. 

\begin{figure}[t!]
    \centering
    \includegraphics[width=.9\columnwidth]{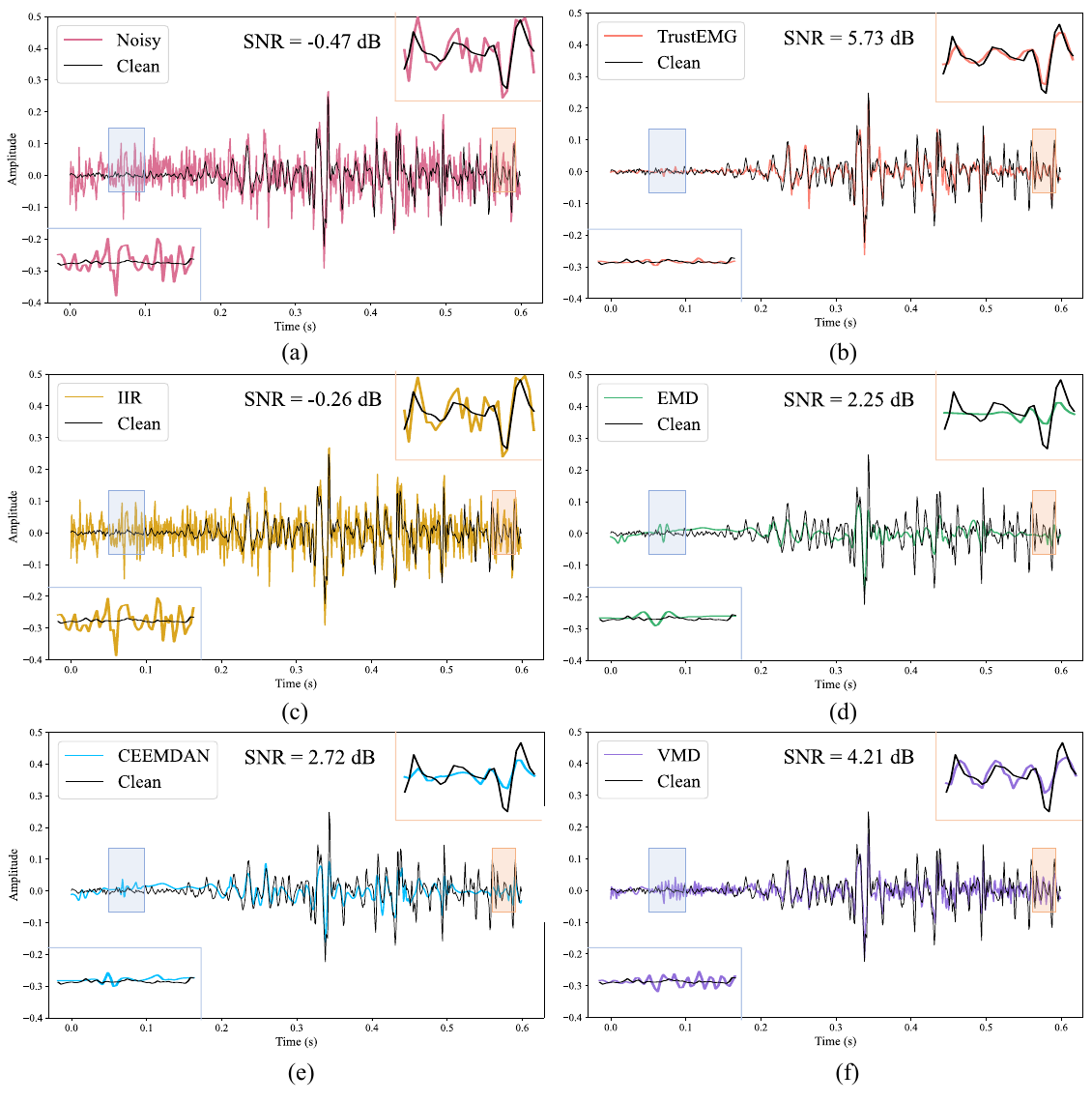}
    \caption{Waveforms of (a) noisy sEMG and enhanced sEMG using (b) TrustEMG-Net, (c) IIR filter, (d) EMD-based method, (e) CEEMDAN-based method, and (f) VMD-based method. The noisy sEMG segment was at SNR of -0.47 dB, extracted from the 2-s noisy sEMG corrupted with WGN at SNR -2 dB. TrustEMG-Net effectively removed contaminants and reconstructed the sEMG, yielding the highest SNR among all the methods.}
    \label{fig:denoised waveform}
\end{figure}

\begin{figure*}[th!]
    \centering
    \includegraphics[width=.68\textwidth]{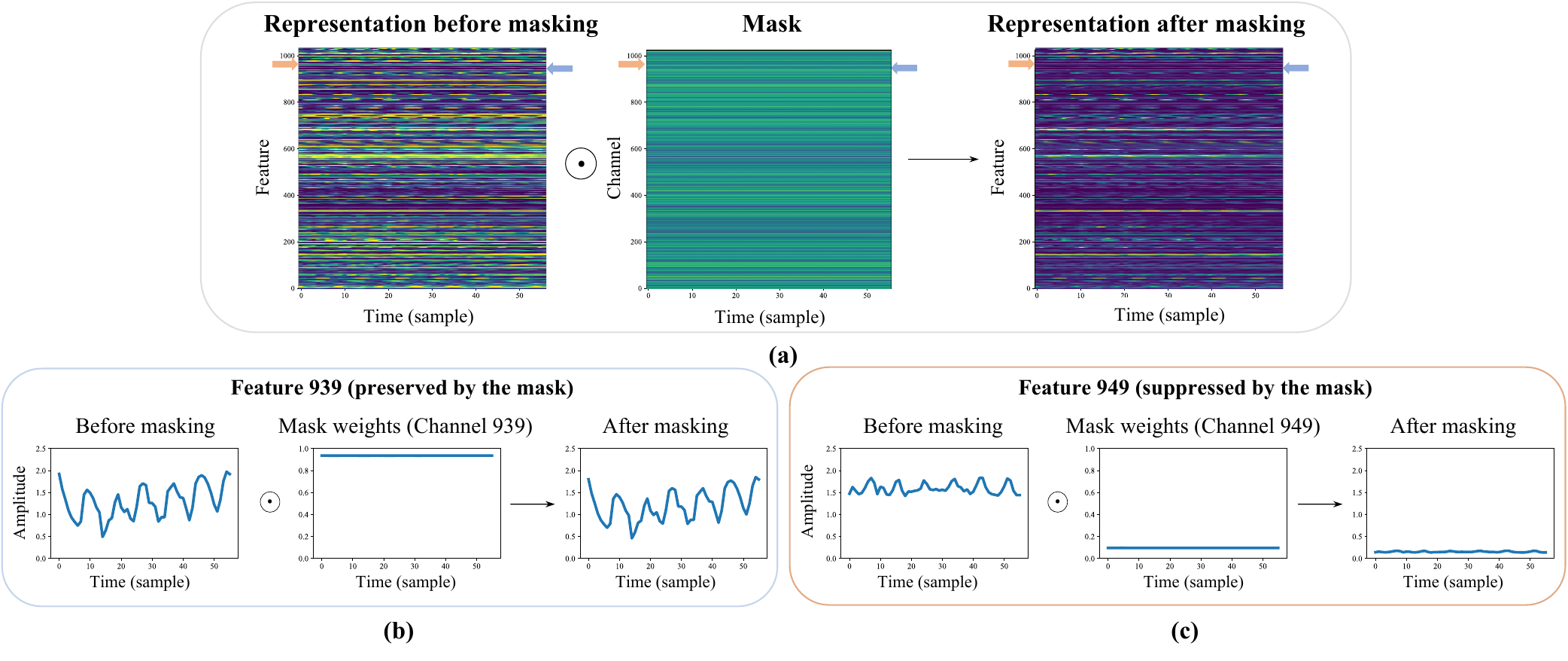}
    \caption{(a) Effect of the RM approach in TrustEMG-Net. The latent representation is derived from input sEMG contaminated by PLI at an SNR of -14 dB. The difference between the latent representation before and after masking indicates that the mask primarily highlights certain feature dimensions of the representation. The mask preserves features by assigning weights close to one, such as (b) feature 939, and it suppresses features by assigning weights close to zero, such as (c) feature 949.}
    \label{fig:mask}
\end{figure*}

\begin{figure*}[th!]
    \centering
    \includegraphics[width=.6\textwidth]{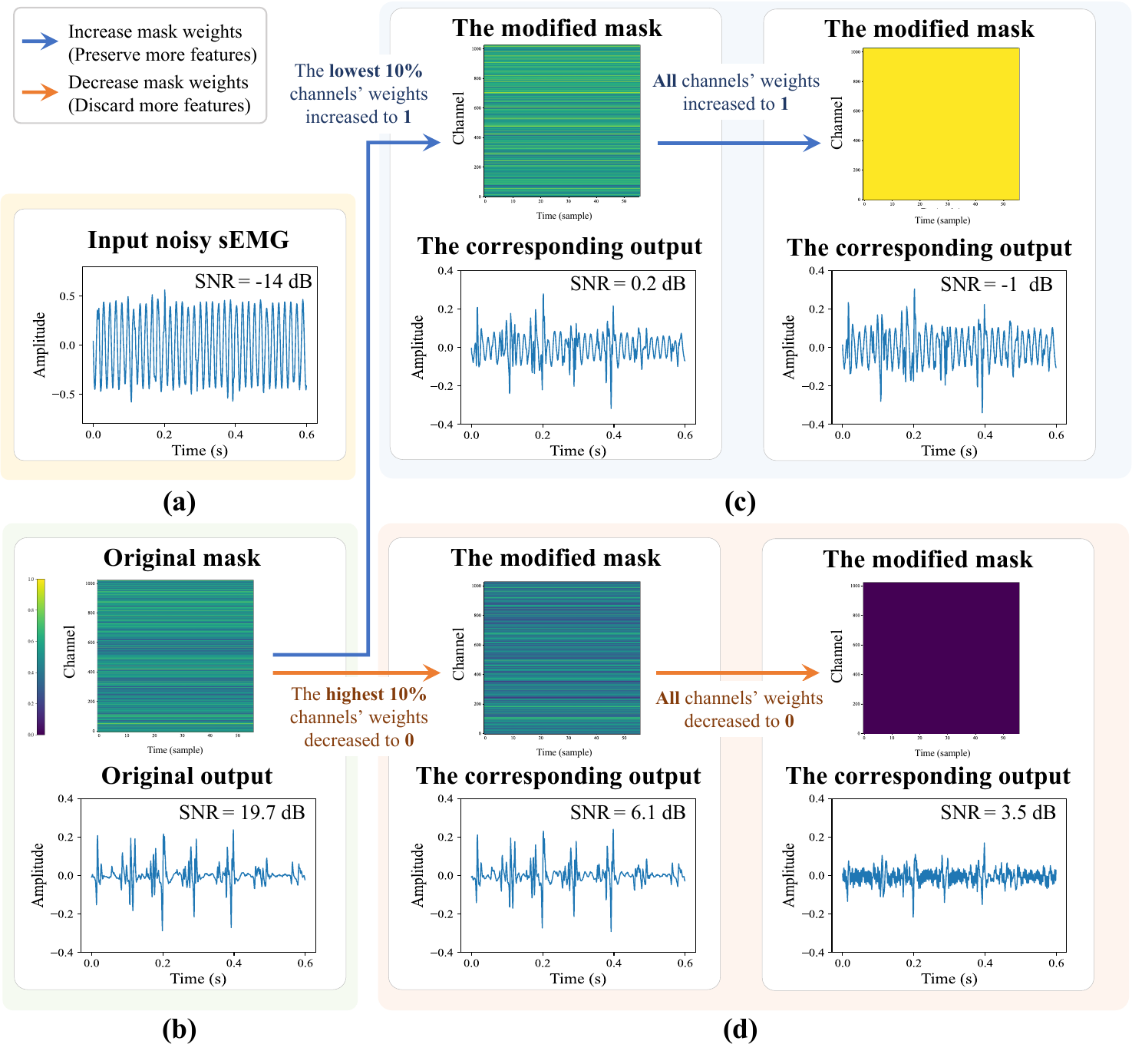}
    \caption{Investigation into the importance of preserving and suppressing specific features in the representation. The observations highlight the importance of optimizing the mask to effectively suppress contaminants while preserving essential sEMG features for optimal denoising results. (a) The input sEMG signal was contaminated by PLI at an SNR of -14 dB. (b) We manually adjusted the weights of the original predicted mask and observed changes in the output waveform. Two modifications are applied: increasing the weights of low channels to 1 (preserving the initially suppressed features) and decreasing the weights of high channels to 0 (suppressing the initially preserved features). (c) As the mask weights increased, the signal quality deteriorated owing to the gradual emergence of PLI, indicating that the features initially suppressed by the mask are closely associated with contaminants. Conversely, (d) as the mask weights increased, the signal quality decreased owing to sEMG distortion, suggesting that preserved features are more related to clean sEMG signals.}
    \label{fig:mask_change_weights}
\end{figure*}

Fig.~\ref{fig:denoised waveform} presents a case of sEMG denoising using each method, demonstrating the superiority of TrustEMG-Net in contaminant removal and high-quality sEMG signal reconstruction compared with other methods. The noisy sEMG segment was contaminated by WGN, which is one of the most challenging contaminant types for sEMG denoising. The results showed that IIR filters had almost no effect on the removal of WGN. The EMD- and CEEMDAN-based methods can remove most contaminants but tend to discard sEMG signals. The VMD-based method performs better in sEMG reconstruction, whereas the contaminant is still visually apparent in the sEMG segments, particularly in the inactive state where no sEMG exists. 

Despite the superiority of TrustEMG-Net across various contaminant types, we observed that existing sEMG denoising methods, including simple IIR filters, outperformed TrustEMG-Net under BW contamination, as shown in Table.~\ref{tab:nonNN_overall}. This outcome can be attributed to the characteristics of BW, which predominantly consists of frequency components below 10 Hz. These frequencies do not overlap with the primary sEMG frequency range (20\textendash 500 Hz). Consequently, unlike the other contaminant types evaluated in this study, BW can be effectively removed without distorting the sEMG using conventional methods. This observation suggests that NN-based methods may not be essential for BW removal when targeting the sEMG frequency range of 20\textendash 500 Hz.

\subsection{Ablation study}
According to our analysis, the outstanding performance of TrustEMG-Net is derived from the U-Net autoencoder structure and the integration of the Transformer encoder using the RM approach.

\subsubsection{U-Net autoencoder structure}
Table~\ref{tab:NN_overall} shows that the U-Net-based methods outperformed CNN and FCN in sEMG contaminant removal under four metrics, except for the RMSE of the MF. The choice of the U-Net structure proves beneficial for the sEMG contaminant removal, which is attributed to the fully convolutional autoencoder and skip connection designs. Convolutional layers have been proven to retain the dependencies within waveform data better than fully connected layers~\cite{FCN_fu2017raw,chiang2019noise}, and skip connections can offer multi-scale features for the decoder to enhance signal reconstruction~\cite{hu2024lightweight}.

We observed that FCN yields better results in the RMSE of the MF compared with TrustEMG-Net under low SNRs (-14 and -10 dB) and three contamination types (BW, ECG, and MOA), with statistically significant differences observed. An interesting finding is that TrustEMG-Net produces a waveform segment with a spectrum that is visually and quantitatively closer to the clean spectrum, but its MF error remains higher than that of FCN. Achieving precise sEMG reconstruction in the frequency or time-frequency domain is crucial for specific clinical or HCI applications~\cite{phinyomark2012usefulness,thongpanja2013mean}. Our training only utilized the L1 loss function on the time-domain waveform and lacked direct optimization for reconstructing the sEMG in the time-frequency domain. This approach does not guarantee optimal performance in metrics involving the frequency domain, such as the MF. The incorporation of a loss function that considers the frequency domain is a potential solution to this problem~\cite{kolbaek2020loss}.

\subsubsection{Representation-masking Transformer}
Integrating the Transformer encoder with U-Net proved to be effective. As shown in Table~\ref{tab:NN_overall},~\ref{tab:NN_different SNR}, and ~\ref{tab:NN_different noise}, both TrustEMG-Net and TrustEMG-Net(DM) achieved higher metric scores than U-Net under most conditions, demonstrating Transformer's ability to model global dependencies within the data. This compensates for the limited perceptive view of convolutional layers, resulting in superior performance compared with the standalone U-Net~\cite{wang2021caunet,wang2021transbts}. 

The RM approach significantly boosts the performance of TrustEMG-Net compared with TrustEMG-Net(DM), particularly in managing narrowband contaminants, such as BW and PLI (Table~\ref{tab:NN_different noise}). This improvement can be attributed to the capability of the masking approach to selectively emphasize the frequency components. 

Fig.~\ref{fig:mask} visually demonstrates how masking impacts the latent representation, which is closely correlated with the spectral information when derived from one-dimensional convolutional layers using the waveform input~\cite{engel2020ddsp}. 
The mask learns to be a spectral feature selector that assigns distinct weights to each feature dimension. If the channel\footnote{Here, "channels" refer to the feature dimensions of the mask, not the electrode channels.} contains high weights close to 1, then the mask preserves the corresponding feature after masking, such as feature 939 in this case. In contrast, if the channel contains low weights close to 0, the mask suppresses the corresponding feature, such as feature 949. This enables the RM to prevent excessive denoising from contaminants that overlap with the sEMG in specific frequency ranges, thereby ensuring robust denoising performance. This RM function is similar to the principles of decomposition-based methods, in which narrowband contaminants typically appear in distinct intrinsic modes and can be effectively filtered~\cite{ma2020emg}.

Fig.~\ref{fig:mask_change_weights} shows the importance of preserving or suppressing specific features in the representation. These findings suggest that features initially suppressed by the mask are closely linked to contaminant components, whereas initially preserved features are more closely associated with sEMG. In this experiment, we manually adjusted the learned mask weights and observed the changes in the output waveform. Specifically, we increased the weights of the low-weight channels to 1 (preserving the initially suppressed features) and decreased the weights of the high-weight channels to 0 (suppressing the initially preserved features). The low- and high-weight channels were determined by ranking the mask averaged along the temporal dimension, which resulted in a 1024-dimensional vector, as the mask primarily functions as a feature selector.

When the mask weights of the low-weight channels were increased to preserve the initially suppressed features, the contaminant (PLI) became apparent in the output waveform, degrading the signal quality. Conversely, reducing the mask weights of the high-weight channels to suppress the initially preserved features resulted in noticeable distortion of the sEMG signal. Hence, optimal denoising requires a mask to selectively suppress and preserve specific features.

\renewcommand{\arraystretch}{.9}
\begin{table}[t]
    \centering
    \caption{Computational complexity of baseline and NN-based sEMG denoising methods.}
    \label{tab:computation}
    \begin{tabular}{ccccc}
     \toprule
     \multirow{2}{*}{Method} & \multirow{2}{*}{\shortstack{Parameter\\number}} & FLOPs & \multicolumn{2}{c}{Computation time (ms)}  \\
     & &  (MAC) & on GPU & on CPU \\
     \midrule
    CNN & 2.69 M & 156.6 M & 8.09 & 14.79\\
    FCN & 11.15 M & 1.41 G & 2.24 & 5.55\\
    U-Net & 16.72 M & 2.36 G & 2.81 & 8.83\\
    TrustEMG-Net(DM) & 25.12 M & 2.84 G & 3.11 & 12.13\\
    TrustEMG-Net & 25.12 M & 2.84 G & 3.55 & 13.14\\
    \midrule
    IIR & - & - & - & 0.33\\
    IIR+TS & - & - & - & 1.20\\
    EMD & - & - & - & 27.27\\
    CEEMDAN & - & - & - & 1263.94\\
    VMD & - & - & - & 1978.39\\
    \bottomrule
    \end{tabular}
\end{table}

\subsection{Computational complexity analysis}
Table~\ref{tab:computation} lists the computational requirements of the baseline and NN-based sEMG denoising methods. The computation time for the baseline methods is specified for the CPU, whereas the NN-based methods include model parameters, floating-point operations (FLOPs), and GPU and CPU computation times. The computation time indicates the duration for processing a 2-s noisy sEMG segment contaminated with a five-type compound contaminant on the Nvidia GeForce RTX 3090 GPU and Intel Xeon Silver 4416+ CPU.

Compared with decomposition-based sEMG denoising methods (EMD, CEEMDAN, and VMD), TrustEMG-Net demonstrates superior efficiency by requiring less computation time, in addition to its effectiveness. This difference largely results from the iterative computational or optimization processes involved in decomposition algorithms. Among NN-based methods, a trade-off exists between the denoising performance and computational efficiency. TrustEMG-Net achieved better denoising results than FCN and U-Net at the cost of more parameters, FLOPs, and computation time. TrustEMG-Net(DM) shares the same parameter count as TrustEMG-Net with a slight increase in computation time owing to the RM technique. Future research could explore techniques such as parameter pruning and quantization~\cite{karnin1990simple, yang2019quantization}, or more efficient sequence-to-sequence models~\cite{gu2023mamba} to mitigate the computational demands for resource-limited or time-sensitive sEMG applications.

\subsection{Limitations and future research}
Although this paper provides insights into NN-based sEMG contaminant removal methods, certain limitations should be acknowledged. First, the experiment included only one sEMG database, the Ninapro database. Future studies should involve additional databases to ensure the generalization capability of the proposed method for different sEMG data. Second, the performance of TrustEMG-Net with real-world noisy sEMG data will be investigated in the future, as the current research relies on synthesized noisy sEMG data. Finally, the proposed approach has not been directly tested in sEMG applications such as hand gesture recognition. Future research should explore the direct effect of denoising on downstream sEMG applications to comprehensively evaluate the practicality and benefits of the proposed denoising algorithm.

\section{Conclusion}
This paper introduces an NN-based denoising method called TrustEMG-Net to effectively remove contaminants from single-channel sEMG signals. The proposed technique incorporates a DAE structure and combines U-Net with a Transformer encoder, utilizing the robust nonlinear mapping capability and data-driven characteristics of NNs to remove sEMG contaminants. Incorporating a representation-masking approach further enhances the denoising capabilities, particularly for narrowband contaminants. The experimental results demonstrated that sEMG signals denoised using TrustEMG-Net exhibit significantly higher signal quality and lower feature extraction errors than existing sEMG denoising methods. This superior performance is consistently maintained across a broad spectrum of SNR inputs and contaminant types. Consequently, the proposed method offers sEMG applications with a more potent, robust, and generalized NN-based approach for contaminant removal.

\section*{Acknowledgment}
We thank Professor Juliano Machado from the Department of Basic, Technical, and Technological Education at Sul-Rio-Grandense Federal Institute for providing us with the MOA dataset used in this research. This work was supported by the Academia Sinica under Grant ASGC-111-M01. 

\section*{References}
\bibliographystyle{IEEEbib}
{\footnotesize
\bibliography{refs}}

\end{document}


\title{Supplementary material}
\maketitle

This is the supplementary material for the paper "TrustEMG-Net: Using Representation-Masking Transformer with U-Net for Surface Electromyography Enhancement." This study compared our proposed method, TrustEMG-Net, with several signal-processing- and neural-network (NN)-based sEMG denoising methods. Sections I and II of this document provide the technical details of these comparison sEMG denoising methods for better understanding and reproducibility. Section III presents an additional experiment for validating the effectiveness of the integration method and representation-masking (RM) approach. 

\section {Signal-processing-based sEMG denoising methods}
This section presents the detailed implementation of the signal-processing-based sEMG denoising methods used in this study, including IIR filters, template subtraction (TS), empirical mode decomposition (EMD), complete ensemble EMD with adaptive noise (CEEMDAN), and variational mode decomposition (VMD).
\renewcommand{\arraystretch}{1.1}
\subsection{IIR filters}
The selection of IIR filters for sEMG contaminant removal is shown in Table~\ref{tab:IIR filters}. Each IIR filter is applied only if the corresponding contaminant exists in sEMG.

\begin{table}[h!]
    \centering
    \caption{IIR filters for sEMG contaminant removal.}
    \begin{tabular}{cc}
         \toprule
         Contaminant & Filter specification \\
         \midrule
         BW &  $4^{th}$ order Butterworth high-pass filter, $f_c$ 10 Hz  \\
         PLI &  Notch filter, $f_c$ 60 Hz and quality factor 5  \\
         MOA~\cite{machado2021deep}&  $4^{th}$ order Butterworth high-pass filter, $f_c$ 40 Hz  \\
         MOA~\cite{moody1984noise}& $4^{th}$ order Butterworth high-pass filter, $f_c$ 20 Hz  \\
         ECG  & $4^{th}$ order Butterworth high-pass filter, $f_c$ 40 Hz  \\
         WGN  & $4^{th}$ order Butterworth band-pass filter, $f_c$ 20 and 500 Hz  \\
         \bottomrule
         \multicolumn{2}{l}{\footnotesize $f_c$ denotes the cutoff frequency.}
    \end{tabular}
    \label{tab:IIR filters}
\end{table}

\subsection{TS+IIR}
TS+IIR applies TS and IIR filters for ECG and non-ECG contaminants, respectively. TS consists of three steps: ECG detection, template extraction, and ECG subtraction. For ECG detection, zero-padding is first performed on signal segments for 0.5 seconds at the front and end of the segments. We then calculate two moving averages (1 s and 0.1 s) and identify ECG-containing segments when these moving averages intersect twice within a specified time (more than 0.14 seconds in this study)~\cite{junior2019template}. Since the sEMG segments are short (2 s), we choose a filtering approach to create ECG templates, using a 4th-order Butterworth high-pass filter with cutoff frequency 50 Hz~\cite{marker2014effects}. After subtracting ECG artifacts, we apply a 4th-order Butterworth high-pass filter with a cutoff frequency of 40 Hz to obtain the best results~\cite{wang2023ecg, drake2006elimination}.

\subsection{EMD-based and CEEMDAN-based methods}
The EMD-based and CEEMDAN-based methods adopt EMD and CEEMDAN for signal decomposition, respectively. The maximum number of IMFs is set to 8 for both methods. As for CEEMDAN, the scale for added noise is set to 0.005, and the number of trials is 20.

The contaminant removal algorithms for both methods refer to previous research~\cite{zhang2016improved, sun2020surface, ma2020emg, naji2011application, sun2023emg}. To discern whether certain contaminant types exist in each mode, we use the Fast Fourier Transform to find the frequency with maximum energy in the spectrum, denoted as $f_{max}$. We then apply the following algorithms for different types of contaminants if included~\cite{ma2020emg}.

\begin{itemize}
\item BW: modes with $f_{max}$ less than 10 Hz are removed. Other modes remove BW by subtracting passing themselves through a 4th-order Butterworth low-pass filter with a cutoff frequency of 10 Hz.
\item PLI: for modes with $f_{max}$ between 50 and 70 Hz, we apply a narrow-band notch filter with center frequency $f_{max}$ and quality factor 20.
\item ECG: we calculate the Pearson correlation coefficients of each mode and ECG template to discern modes with more ECG contamination. The ECG template here is extracted from noisy sEMG by applying a 4th-order Butterworth high-pass filter with a cutoff frequency of 40 Hz. For the five modes with higher correlations with the ECG template, we pass them through a 4th-order Butterworth high-pass filter with a cutoff frequency of 30 Hz.
\item MOA: modes with $f_{max}$ less than 20 Hz are removed. Other modes are passed through a 4th-order Butterworth high-pass filter with a cutoff frequency of 20 Hz.
\item WGN: the first mode is removed. For other modes, whether to remove them is determined based on their standard deviation. If the standard deviation of the mode exceeds the corresponding threshold value, we apply wavelet thresholding to the mode with sym8 as the mother wavelet~\cite{zhang2016improved, sun2020surface, sun2023emg}.
\end{itemize}

\subsection{VMD-based method}
The VMD-based method conducts signal decomposition using VMD with a parameter set as follows. The number of IMF is 10, the penalty factor is 1000, and the tolerance is 0.001~\cite{ma2020emg}. The initialization of center frequency adopts uniform distribution.

We follow~\cite{ma2020emg, ashraf2023variational} and apply the following algorithms for different contaminants. The usage of $f_{max}$ is identical to the CEEMDAN-based methods. Notably, we directly apply the same IIR filter to eliminate ECG artifacts, as there is currently no VMD-based method for ECG contamination.
\begin{itemize}
\item BW, PLI, and MOA: identical to the CEEMDAN-based method.
\item WGN: the soft iterative interval thresholding method is applied, and the parameter refers to ~\cite{ashraf2023variational}. The only difference is that the threshold value is divided by 4 to yield better denoising results in this study.
\end{itemize}

\section{NN-based sEMG denoising methods}

This section demonstrates the structures of NN models used for the ablation study of TrustEMG-Net, including CNN, FCN, and U-Net. Note that all the kernels of the convolutional layers of these models are 1-dimensional.

\subsection{CNN}

The CNN consists of four convolutional layers and two fully connected layers. Except for the last fully connected layer, which is a linear layer, all other layers use the ReLU activation function and batch normalization. The parameters for the convolutional layers are shown in Table~\ref{tab:conv_layers}. 

\renewcommand{\arraystretch}{1.2}
\begin{table}[h!]
    \centering
    \scriptsize
    \caption{Parameters of the convolutional layers in CNN.}
     \label{tab:conv_layers}
    \begin{tabular}{ccccc}
        \toprule
        Layer & Input channel & Filter number & Filter size & Stride \\
        \midrule
        Convolutional layer 1 & 1 & 16 & 8 & 2 \\
        Convolutional layer 2 & 16 & 32 & 8 & 2 \\
        Convolutional layer 3 & 32 & 64 & 8 & 2 \\
        Convolutional layer 4 & 64 & 128 & 8 & 2 \\
        \bottomrule
    \end{tabular}
\end{table}
The dimensions of the two fully connected layers are 400 and 200, respectively. Due to the large number of parameters in the first fully connected layer, dropout regularization with a dropout rate of 50\% 

Note that the input sEMG signal is segmented into 200 points per segment, as we discover that smaller segments can yield better results on CNN. However, segmentation increases the computation time of the CNN. All the outputs are concatenated after passing through the CNN to form the denoised signal. 

\subsection{FCN}
The FCN's encoder and decoder each consist of five convolutional layers and five transposed convolutional layers, with detailed parameters shown below. Each layer uses the ReLU activation function and batch normalization except for the last transposed convolutional layer.

\subsection{U-Net}
The architecture of the U-Net is the same as the structure of TrustEMG-Net, as shown in Table~\ref{tab:conv_transpose_layers}. The only difference is that the bottleneck of the U-Net does not employ a Transformer encoder.

\begin{table}[]
    \centering
    \scriptsize
    \caption{Parameters of the convolutional layers in FCN.}
    \label{tab:conv_transpose_layers}
    \begin{tabular}{ccccc}
        \toprule
        Layer & Input channel & Filter number & Filter size & Stride \\
        \midrule
        Convolutional layer 1 & 1 & 64 & 8 & 2 \\
        Convolutional layer 2 & 64 & 128 & 8 & 2 \\
        Convolutional layer 3 & 128 & 256 & 8 & 2 \\
        Convolutional layer 4 & 256 & 512 & 8 & 2 \\
        Convolutional layer 5 & 512 & 1024 & 8 & 2 \\
        ConvTranspose layer 1 & 1024 & 512 & 8 & 2 \\
        ConvTranspose layer 2 & 512 & 256 & 8 & 2 \\
        ConvTranspose layer 3 & 256 & 128 & 8 & 2 \\
        ConvTranspose layer 4 & 128 & 64 & 8 & 2 \\
        ConvTranspose layer 5 & 64 & 1 & 8 & 2 \\
        \bottomrule
    \end{tabular}

\end{table}

\begin{figure*}[t]
\centering
    \includegraphics[width=.9\textwidth]{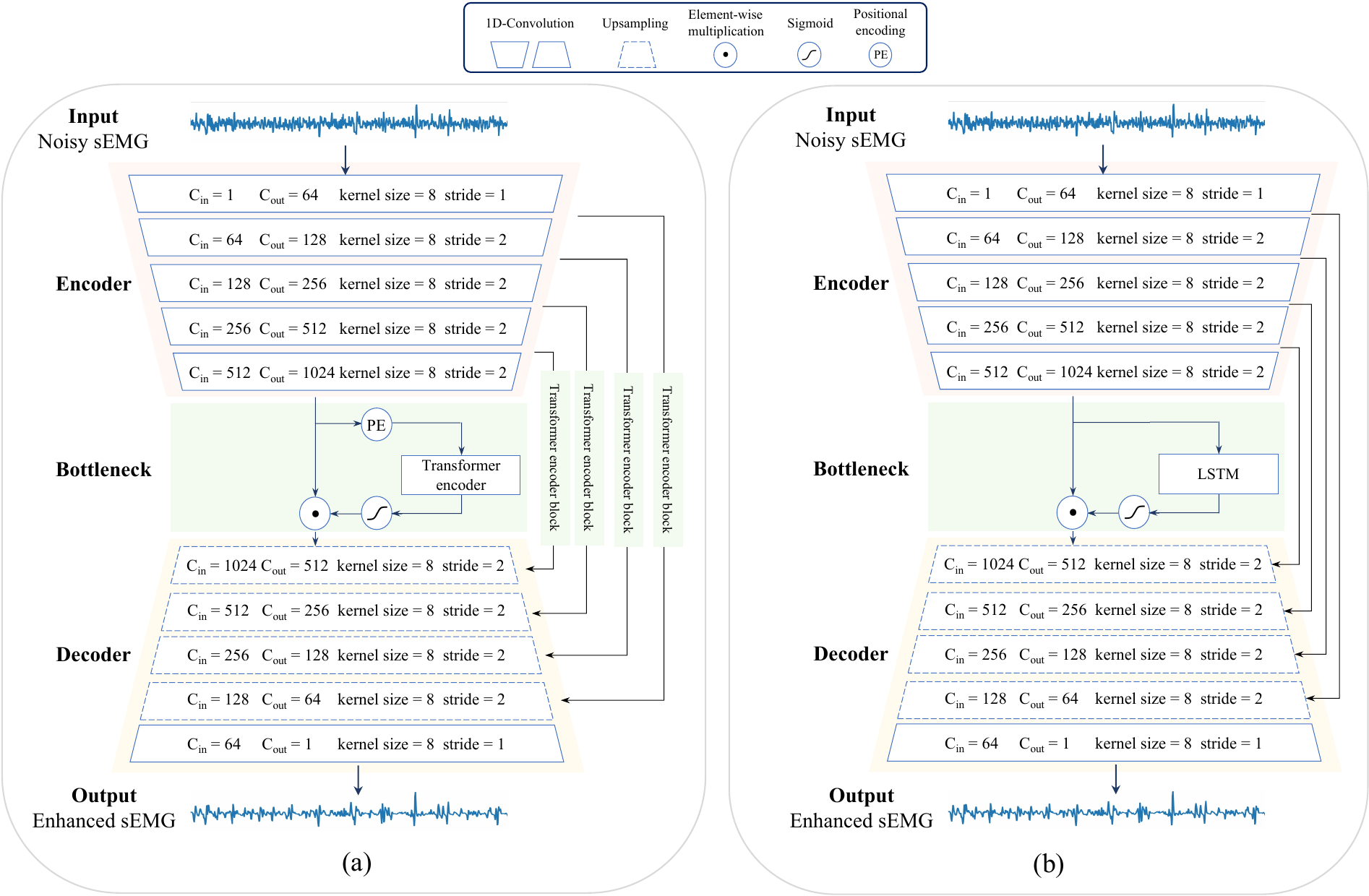}
    \caption{The architecture of (a) U-Net+5Trans and (b) U-Net+LSTM.}
    \label{fig:integrate}
\end{figure*}

\renewcommand{\arraystretch}{1.5}
\begin{table*}[h!]
    \centering
    \caption{Overall performance of NN models with different integration methods for sEMG contaminants removal.}
    \label{tab:integrate_overall}
    \begin{tabular}{ccccccc}
    \toprule
    Metrics & U-Net+5Trans(DM) & U-Net+5Trans & U-Net+LSTM(DM) & U-Net+LSTM & TrustEMG-Net(DM) & TrustEMG-Net \\
    \cmidrule{2-7}
    SNR$_{imp}$ (dB) & 11.80 $\pm$ 0.51* &13.42 $\pm$ 0.47* &13.34 $\pm$ 0.41* & 13.46 $\pm$ 0.46* &13.45 $\pm$ 0.45* &\bf 13.64 $\pm$ 0.38  \\ 
    RMSE ($\times 10^{-2}$) & 2.50 $\pm$ 0.34* &2.22 $\pm$ 0.30* &2.24 $\pm$ 0.30* &2.22 $\pm$ 0.31* &2.21 $\pm$ 0.30* &\bf 2.18 $\pm$ 0.28  \\
    PRD (\%) & 56.29 $\pm$ 2.76* &49.32 $\pm$ 2.03* &49.57 $\pm$ 1.91* &49.11 $\pm$ 1.94* &49.05 $\pm$ 2.00* &\bf 48.44 $\pm$ 1.70  \\
    RMSE of ARV ($\times 10^{-3}$) & 11.89 $\pm$ 2.12* &9.16 $\pm$ 1.70* &8.98 $\pm$ 1.63* &8.93 $\pm$ 1.62* &8.90 $\pm$ 1.58* &\bf 8.72 $\pm$ 1.52  \\
    RMSE of MF (Hz) & 18.36 $\pm$ 1.86* &16.84 $\pm$ 0.81 &17.78 $\pm$ 0.98* &17.63 $\pm$ 1.16* & 16.94 $\pm$ 0.94* &\bf 16.63 $\pm$ 0.91 \\
    \bottomrule
    \multicolumn{7}{l}{\footnotesize *Denotes a significant difference (\textit{p}-value $<$ 0.05) with the proposed method. \textbf{Bold} font indicates the best score for each metric.}
    \end{tabular}
\end{table*}

\section{Comparison with other integration methods}
This section compares TrustEMG-Net with two other integration approaches featuring similar model architectures. The first integration method, U-Net+5Trans, integrates five one-layer Transformer encoders into all four skip connections and the bottleneck of U-Net~\cite{yun2021spectr}, as shown in Fig.~\ref{fig:integrate} (a). The Transformer encoders at the skip connections follow TrustEMG-Net's settings, with adjustments made to the embedding dimensions to match the feature dimension of each skip connection. The second integration method, U-Net+LSTM, replaces TrustEMG-Net's Transformer encoder with a one-layer Long Short-Term Memory (LSTM) model~\cite{he2020automatic}, which is also known for its ability to capture long-term temporal dependencies~\cite{hochreiter1997long}. The structure of U-Net+LSTM is shown in Fig.~\ref{fig:integrate} (b). Additionally, both integration methods incorporate direct mapping (DM) and representation masking (RM) approaches, resulting in four distinct model structures. Models using DM are denoted by "(DM)" following their model types.


\subsection{Comparison between integration methods}
Table~\ref{tab:integrate_overall} presents the overall performance comparison between TrustEMG-Net and other integration methods. It is evident that TrustEMG-Net consistently outperforms all other integration approaches across all five evaluation metrics, with statistically significant differences except when compared with U-Net+5Trans in terms of RMSE of MF. This underscores the effectiveness of TrustEMG-Net's integration approach, which embeds the Transformer encoder at the bottleneck of the U-Net architecture.

We observe that the denoising performance of U-Net+5Trans does not benefit from incorporating more Transformer encoders at skip connections, which can lead to overfitting in this task. Moreover, TrustEMG-Net performs better than U-Net+LSTM. We attribute this outperformance to the parallel processing characteristics of the Transformer encoder, enabling TrustEMG-Net to exhibit better global information extraction capability and computational efficiency compared with U-Net+LSTM.

\subsection{Effectiveness of the RM approach}
Table~\ref{tab:integrate_overall} also highlights the effectiveness of the RM approach in sEMG contaminant removal. Across various integration frameworks (i.e., U-Net+5Trans and U-Net+LSTM), RM-based versions consistently achieve superior denoising results compared with their DM-based counterparts, with all performance differences being statistically significant (\textit{p}-values = 0.000 and 0.0026, respectively.)


\section*{References}
\bibliographystyle{IEEEbib}
{\footnotesize
\bibliography{refs}}